\documentclass{iopart}
\usepackage{iopams}
\usepackage{amssymb}
\usepackage{graphicx}
\usepackage{color}
\usepackage{ulem}

\def\a{\alpha}
\def\b{\beta}

\def\m{\mu}
\def\n{\nu}

\def\l{\lambda}

\def\d{\delta}

\def\be{\begin{equation}}
\def\ee{\end{equation}}
\def\bea{\begin{eqnarray}}
\def\eea{\end{eqnarray}}
\def\nn{\nonumber}

\begin{document}

\title{Black holes in Lorentz-violating gravity theories}

\author{Enrico Barausse$^1$ and Thomas P. Sotiriou$^2$}

\address{$^1$Institut d'Astrophysique de Paris, UMR 7095 du CNRS,
Universit\'{e} Pierre \& Marie Curie, 98bis Bvd. Arago, 75014 Paris, France\\
$^2$SISSA, Via Bonomea 265, 34136, Trieste, Italy {\rm and} INFN, Sezione di Trieste, Italy}

\begin{abstract}
Lorentz-symmetry and the notion of light cones play a central role in the definition of horizons and the existence of black holes. Current observations provide strong indications that  astrophysical black holes do exist in Nature. Here we explore what happens to the notion of a black hole in gravity theories where local Lorentz symmetry is violated, and discuss the relevant astrophysical implications. Einstein-aether theory and Ho\v rava gravity are used as the theoretical background for addressing this question. We review earlier results about static, spherically symmetric black holes, which demonstrate that in Lorentz-violating theories there can be a new type of horizon and, hence, a new notion of  black hole. We also present both known and new results on slowly rotating black holes in these theories, which provide insights on how generic these new horizons are. Finally, we discuss the differences between black holes in Lorentz-violating theories 
and in General Relativity, and assess to what extent they can be probed with present and future observations.
\end{abstract}

\pacs{04.50.Kd, 04.70.Bw}

\section{Introduction}
\label{intro}

Einstein's equations admit black hole solutions, and gravitational collapse in General Relativity inevitably leads to the formation of
these objects, as shown by the celebrated singularity theorems ~\cite{Penrose:1964wq,Hawking:1969sw}.  Observational evidence for the existence of black holes is 
strong as well, both in the ``stellar-mass'' range ($M\sim 5$ -- $20 M_\odot$), and in the ``supermassive'' range ($M\sim 10^5$ -- $10^9 M_\odot$). 
In the stellar-mass range, black-hole candidates in X-ray binaries are too heavy to be neutron stars or quark stars with reasonable equations of state~\cite{ruffini,kalogera}; in the supermassive range,  observations of stars around the black-hole candidate in the center of the Milky Way (SgrA$^*$) show
that this object is too compact and massive to be a dark cluster of low-luminosity bodies~\cite{darkcluster} or a fermion star~\cite{schoedel}. Also, near-infrared observations
of SgrA$^*$ seem to provide circumstantial evidence that this object has an event horizon (because if it had a surface, it would emit radiation, {\it e.g.}~thermally, 
and be more luminous than observed)~\cite{narayan_horizon}, thus disfavoring more exotic horizon-less alternatives such as boson stars or gravastars. 
Similar observations also exist in low-mass X-ray binaries for stellar-mass black-hole candidates,
which do not show the type-I bursts that would be expected from the interaction of the accreting gas with their surface, 
if they had one~\cite{typeIbursts1,typeIbursts2,typeIbursts3}. Moreover, the 
current paradigms for quasars and active galactic nuclei, as well as for the formation of galaxies,
also require that supermassive black holes be present in the center of almost every galaxy. In fact, stellar and gas dynamical evidence~\cite{kormendy} 
and observations of the optical, ultraviolet (UV) and X-ray spectrum of active galactic nuclei~\cite{reverberation_mapping}
do indeed corroborate this assumption.

The defining characteristic of a black hole is, as mentioned above, the presence of an event horizon. That 
is a causal boundary that separates (or defines) the interior and exterior of a black hole, and 
prevents any signal originating from the interior from reaching the exterior.
The reason why spacetimes that posses an event horizon can exist in General Relativity is rooted in the causal structure
of the theory. The latter is inherited from Special Relativity, and 
confines signals, irrespective of their nature, to propagate within future-directed light cones.
Although event horizons depend in general on the global structure of the spacetime, in static, spherically symmetric spacetimes (and in stationary, slowly rotating ones, which we will also consider here) they are usually discussed in terms of the structure of light cones in a well-behaved coordinate system, i.e.~they correspond to the location where these light cones ``tilt'' and cease allowing signals to propagate ``outwards''.
It is, therefore, clear that local Lorentz invariance, being the symmetry underpinning Special and General Relativity, 
is central to the existence of black holes.
It is then natural to wonder what happens to horizons and black holes if Lorentz invariance is not an exact symmetry of Nature.

Intuitively, the answer to this question seems to depend on the way local Lorentz symmetry is violated, and on how this affects the causal structure
of the theory. Consider for instance a theory with a preferred frame, but where all excitations have linear dispersion relations. 
Lorentz symmetry is then broken by the existence of a preferred frame,
and certain modes may perhaps propagate faster than light in that frame,
but all propagating modes have a finite speed. Therefore,
there are still ``propagation cones'' in such a Lorentz-violating theory, and the causal structure remains qualitatively similar to General Relativity. 
The modification that one may expect at the level of black holes  is the presence of multiple horizons, i.e. modes propagating at different speeds would have different causal boundaries.

However, there is no particular reason why dispersion relations should be linear once Lorentz symmetry has been given up. Consider instead dispersion relations 
\be
\label{mdsr}
\omega^2\propto k^2+\alpha k^4+\ldots\, ,
\ee
where $\omega$ is the frequency, $k$ the wave-number, $\alpha$ a constant with appropriate dimensions, and the dispersion relation can potentially include higher powers of $k$. Perturbations with sufficiently large $k$ can then travel arbitrarily fast in the preferred frame. Therefore, there are no propagation cones, and the causal structure is drastically different from Special and General Relativity, as the existence of a preferred frame and
 infinite-speed propagation imply the existence of hypersurfaces of simultaneity. The very notion of a black hole does not seem to fit in this framework. 

It is also worth mentioning that if one were to start with a theory with a higher-order dispersion relation, and then truncate it 
to order $k^2$ as a result of some low-momentum approximation, 
then the picture would drastically change into the one depicted  above for 
a linear dispersion relation. So, one could say that excitations with sufficiently low momenta experience 
the existence of an event horizons, but more energetic ones would be able to escape them. Alternatively, one could say that the existence of horizons is a low-energy artefact. 

This conclusion is obviously based on highly heuristic arguments and lacks a firm mathematical basis. It does, however, act as the motivation and starting point 
for a rigorous exploration of what happens to the notion of a black hole in gravitational theories that exhibit violations of Lorentz invariance.
This is indeed the purpose of this paper. We stress that this is not a purely academic question. As discussed above, 
the astrophysical evidence for the existence of black holes is rather convincing, and observations allow us to determine the 
characteristics of objects with ever-increasing accuracy. Hence, black holes can be powerful laboratories for gravity, 
and in particular they can be used to constrain violations of local Lorentz symmetry in the gravity sector. 
Such constraints would be complementary to the very stringent constraints already existing for Lorentz-symmetry violations in the matter sector~\cite{Kostelecky:2008ts,Liberati:2013xla}. 

Quantifying the potential violations of symmetries that are considered as fundamental is an important goal in itself, but
it requires a theoretical framework providing predictions to be confronted with observations. Einstein-aether theory (\ae-theory) \cite{Jacobson:2000xp} 
has been quite successful at playing this role when it comes to Lorentz-invariance violations in gravity. 
\AE-theory is essentially General Relativity coupled to a vector field (the aether), which is constrained to have unit norm and be timelike.
Being of unit length, the aether cannot vanish, and thus breaks boost invariance by defining (locally) a preferred frame. 
Recently, additional motivation for Lorentz-violating gravity theories has been put forward:
Ho\v rava has suggested that giving up Lorentz symmetry can lead to considerably better UV 
properties, and it appears possible to put together a power-counting renormalizable Lorentz-violating theory of gravity~\cite{arXiv:0901.3775}. 
This could be a pay-off significant enough to entertain the idea that Lorentz symmetry might indeed be approximate, and not exact, in the gravity sector.

We will give a brief overview of both \ae-theory and Ho\v rava gravity in section \ref{framework}. 
In connection with what we discussed above, it is worth pointing out that dispersion relations are linear in the former,
and higher-order in the latter.
Moreover, the low-energy limit of Ho\v rava gravity, which is a truncation of the theory at second order in derivatives, 
has been shown to be dynamically equivalent to \ae-theory, provided that 
the aether is forced to be hypersurface-orthogonal at the level of the action \cite{Jacobson:2010mx}. 
Together, the two theories form an excellent theoretical framework for rigorously exploring the fate of black holes once local Lorentz symmetry is violated. 
In section \ref{sphBH} we indeed investigate static, spherically symmetric, asymptotically flat black holes in the two theories, 
following the lines of Ref.~\cite{Barausse:2011pu}, while in section \ref{slowly_rotating} we move away from spherical symmetry and focus
on slowly-rotating black holes (partially following Refs.~\cite{letter,arxiv_note} but providing independent derivations
of the results of those papers, and clarifying their physical interpretation). Section \ref{cons} contains our conclusions.

 In this paper, we use a metric signature $(+---)$ and units where $c=\hbar=1$.

\section{Framework}
\label{framework}

\subsection{\AE-theory and Ho\v rava gravity}
\label{aeandhl}

The action for \ae-theory is
\be \label{S}
S_{\ae} = \frac{1}{16\pi G_{\ae}}\int \sqrt{-g}~ (-R -M^{\a\b}{}_{\m\n} \nabla_\a u^\m \nabla_\b u^\n)
~d^{4}x \ee
where $R$ is the Ricci scalar, $g$ the determinant of the metric $g_{ab}$, and
\be M^{\a\b}{}_{\m\n} = c_1 g^{\a\b}g_{\m\n}+c_2\d^{\a}_{\m}\d^{\b}_{\n}
+c_3 \d^{\a}_{\n}\d^{\b}_{\m}+c_4 u^\a u^\b g_{\m\n}\,,
\ee
with $\nabla_\m$ being the covariant derivative associated to $g_{\mu\nu}$ and $c_i$ being dimensionless coupling constants. 
Also, $u^\m$ is constrained to be a unit timelike
vector, i.e. 
\be
g_{\m\n}u^\m u^\n=1\,.
\ee
This can be explicitly imposed 
by adding the Lagrange multiplier term $\l(g_{\m\n}u^\m u^\n-1)$ to the action, or it can be taken into account implicitly during the variation.
This constraint is a key feature of \ae-theory: by virtue of it, the aether can never vanish, and always points to some timelike direction. 
So, in the locally flat coordinate system where $g_{\mu\nu}$ becomes the Minkowski metric, 
the aether breaks boost invariance, leading to violations of Lorentz symmetry and defining a preferred frame. 

Consider now a restricted version of \ae-theory where the aether is always hypersurface-orthogonal. Then, one can express the aether in terms of a scalar $T$ as
\be
\label{ho}
u_\a=\frac{\partial_\a T}{\sqrt{g^{\m\n}\partial_\m T \partial_\n T}}\,,
\ee
where the unit constraint has already been taken into account. Because the aether must be timelike, $T$ defines a time coordinate, and one can choose to foliate spacetime into 
constant-$T$ hypersurfaces. In this adapted foliation, one has
\be\label{preferred foliation}
u_\a=\delta_{\a}^{T} (g^{TT})^{-1/2}=N\delta_{\a}^{T}  \,,
\ee
where $N=(g^{TT})^{-1/2}$ is the lapse function, and the action takes the form
\be\label{SBPSH}
S_{h.o.\ae}=\frac{1}{16\pi G_{H}}\!\int dT d^3x \, N\sqrt{h} \, (K_{ij}K^{ij} - \lambda K^2 
+ \xi {}^{(3)}\!R + \eta a_ia^i)\,,
\ee
where  $K^{ij}$ is the extrinsic curvature of the constant-$T$ hypersurfaces,  $h_{ij}$ is the induced 
 spatial
 metric, ${}^{(3)}\!R$ its Ricci curvature, and
\be
a_i=\partial_i \ln{N}\,
\ee
is the
acceleration of the aether flow. We have also introduced new parameters which are given in terms of those appearing in action (\ref{S}) as
\be
\label{HLpar}
\frac{G_H}{G_{\ae}}=\xi=\frac{1}{1-c_{13}}, \quad \lambda=\frac{1+c_2}{1-c_{13}},\quad \eta=\frac{c_{14}}{1-c_{13}},
\ee
where $c_{ij}=c_i+c_j$. Only three coupling constants appear in action (\ref{SBPSH}), as opposed to the four $c_i$ in action (\ref{S}). This is because for a hypersurface-orthogonal aether field,
 any one of the $c_1$, $c_3$ or $c_4$ terms can be expressed in terms of the other two without loss of generality.

Action (\ref{SBPSH}) is what is referred to as the low-energy (or infrared) limit of Ho\v rava gravity. In its complete version the action of Ho\v rava gravity is \cite{Blas:2009qj}
\be
\label{SBPSHfull}
S_{HL}= \frac{1}{16\pi G_{H}}\int dT d^3x \, N\sqrt{h}\left(L_2+\frac{1}{M_\star^2}L_4+\frac{1}{M_\star^4}L_6\right)\,,
\ee
where 
\be\label{L2}
L_2
=K_{ij}K^{ij} - \lambda K^2 
+ \xi {}^{(3)}\!R + \eta a_ia^i,
\ee
$M_\star$ is a new mass scale, 
and $L_4$ and $L_6$ are respectively of fourth and sixth order in the spatial derivatives, but contain no derivatives with respect to $T$. 
The presence of sixth-order terms in the spatial derivatives is crucial for power-counting renormalizability \cite{arXiv:0901.3775}. 
A more precise prescription to construct $L_4$ and $L_6$ follows from the symmetries of the low-energy part of the action. Action (\ref{SBPSH}) is written in a preferred foliation of $T=$ constant hypersurfaces, and the existence of this foliation is essential for the addition of the $L_4$ and $L_6$ terms. The action is invariant under diffeomorphisms that preserve this foliation, $T\rightarrow T'(T)$ and $x^i\rightarrow x'^i(x^i,T)$, but not under arbitrary diffeomorphisms. Therefore, $L_4$ and $L_6$ should contain (respectively) all forth and sixth-order
terms that are compatible with this symmetry. In practice, this means that $L_4$ and $L_6$ can contain any 3-dimensional scalar constructed with  $a_i$, $h_{ij}$ and their (spatial) derivatives.

We have deliberately chosen to introduce Ho\v rava gravity through its relation with \ae-theory for various reasons (see also Ref.~\cite{Blas:2009yd}). First, the relation between the two theories will be extensively used in what follows. Second, 
this choice shows that Ho\v rava gravity can be written in a covariant manner. Third, it makes it straightforward to argue that the theory has an extra scalar degree of freedom 
(since the transverse mode of the aether disappears due to the hypersurface-orthogonality condition), which becomes apparent 
and defines the preferred foliation. Note that different versions of Ho\v rava gravity exist with extra symmetries, or different field content \cite{Sotiriou:2009bx, Weinfurtner:2010hz,Horava:2010zj,Vernieri:2011aa,Vernieri:2012ms}, 
but we will not consider them here (see Ref.~\cite{Sotiriou:2010wn} for a brief review).

The presence of terms of fourth and sixth order in the spatial derivatives  leads to higher-order dispersion relations for the spin-2 graviton and the scalar degree of freedom, 
such as those discussed in the Introduction. In \ae-theory instead, all degrees of freedom have linear dispersion relations. The same property is shared by the low-energy 
limit of Ho\v rava gravity, 
but is an artefact of the truncation. Thus, in combination, the two theories do indeed act as an excellent framework to address the issues raised in the Introduction. 
 
The precise form of $L_4$ and $L_6$ will {\it not} be crucial for what follows. Finding black hole solutions when those terms are
included is very challenging from a computational and numerical point of view, so here we will rather focus on \ae-theory and the low-energy part of Ho\v rava gravity.
In particular, we will look for spacetimes with a metric horizon. This is a null hypersurface of the covariant
4-dimensional metric (which couples minimally to the matter fields, as required by the weak equivalence principle), and acts as causal horizon for those fields. Besides the metric horizon, these solutions are expected to possess different causal horizons for the propagating modes that reside in the gravity sector (i.e.~the spin-2 and spin-0 modes of Ho\v rava gravity, and the
spin-2, spin-1 and spin-0 modes of \ae-theory), because these modes will generically travel at superluminal or subluminal speeds. These horizons will be 
null surfaces of the effective metrics $g^{(i)}_{\a\b}=g_{\a\b}+(s_i^2-1)u_\a u_\b$
where $s_i$ is the speed of the spin-$i$ mode (see Ref.~\cite{Eling:2006ec} for a  detailed discussion).\footnote{The instantaneous mode of Ho\v rava gravity is  special in this respect \cite{Blas:2011ni}.}
We will then proceed and discuss the effect the higher-order derivative terms $L_4$ and $L_6$ on the causal structure of these spacetimes, and discuss whether they really deserve to 
be called ``black holes''.

Our analysis hinges on the assumption that $M_\star$ is a sufficiently large mass scale, so that the presence of the $L_4$ and $L_6$ terms does not really affect the spacetime geometry 
 in regions of interest for our analysis. This will be discussed in detail in the next section. $M_\star$ is indeed bound from below from tests of Lorentz symmetry. In particular, purely gravitational experiments impose the mild bound $M_{\star}\gtrsim 10^{-3}$ eV. Test of Lorentz symmetry in the matter sector can lead to constraints that are many orders of magnitude higher for $M_\star$, but which depend strongly on the details of the percolations of Lorentz-symmetry breaking from the gravity to the matter sector \cite{Burgess:2002tb,Iengo:2009ix,Pospelov:2010mp,Blas:2010hb,Liberati:2012jf}. $M_\star$ is also, perhaps more surprisingly, bound from above
($M_{\star}\lesssim 10^{16}$ GeV), from the requirement that the theory behave perturbatively at all scales \cite{Papazoglou:2009fj,Kimpton:2010xi,Blas:2009ck}, so that the power-counting renormalizability arguments  put forth in Ref.~\cite{arXiv:0901.3775} actually apply.

We conclude this brief overview of \ae-theory, Ho\v rava gravity and their mutual relation with a brief discussion of their field equations. 
Varying the action (\ref{S}) with respect to the metric $g^{\a\b}$ and the aether $u^\mu$, in the absence of matter fields and taking into account the unit constraint, yields 
\bea\label{gfe}
E_{\a\b}&\equiv &  -\frac{1}{2} ( G_{\a\b} - T^{\ae}_{\a\b})=0\,,\\
 \label{AEeq}
\AE_\mu& \equiv &\left(\nabla_\a J^{\a\n}-c_4\dot{u}_\a\nabla^\n u^\a\right) \left(g_{\mu\nu}-u_\m u_\n\right)=0
\eea
where $G_{\a\b}=R_{\a\b}-R g_{\a\b}/2$ 
is the  Einstein tensor,
\bea\label{Tae}
T^{\ae}_{\a\b}&=&\nabla_\m\left(J^{\phantom{(\a}\m}_{(\a}u_{\b)}-J^\m_{\phantom{\m}(\a}u_{\b)}-J_{(\a\b)}u^\m\right)\nonumber\\
&&+c_1\,\left[ (\nabla_\m u_\a)(\nabla^\m u_\b)-(\nabla_\a u_\m)(\nabla_\b u^\m) \right]\nonumber\\
&&+\left[ u_\n(\nabla_\m J^{\m\n})-c_4 \dot{u}^2 \right] u_\a u_\b
+c_4 \dot{u}_\a \dot{u}_\b-\frac{1}{2} L_{\ae} g_{\a\b}\,,
\eea
is the aether stress-energy tensor, and
\be
J^\a_{\phantom{a}\m}=M^{\a\b}_{\phantom{ab}\m\n} \nabla_\b u^\n\,
\qquad\dot{u}_\n=u^\m\nabla_\m u_\n\,.
\ee
Taking instead the variation with respect to $g^{\a\b}$ and $u_\mu$ one obtains, again in the absence of matter fields and taking into account the unit constraint,
\bea
E_{\a\b}+\AE_{(\a}u_{\b)}=0\,,\label{EinsteinAE}\\ \AE_\mu=0\label{AEeq2}\,.
\eea
These equations are clearly equivalent to the system~\eref{gfe} and \eref{AEeq}, and either set of equations represents the full field equations
of \ae-theory. (We will use both forms of the field equations in what follows).

Imposing the constraint \eref{ho} before the variation, and then varying the action (\ref{S}) with respect to $g_{\a\b}$ and $T$ yields
\bea\label{hl1}
&&E_{\a\b}+\AE_{(\a}u_{\b)}=0\,,\\
 \label{hleq}
&&\nabla_\mu \left(\frac{\AE^\mu}{\sqrt{\nabla^\alpha T \nabla_\alpha T}}  \right)=0\,,
\eea
keeping the same notation as above. These are the equations of the low-energy limit of Ho\v rava gravity in covariant form. 

Based on the above, it is  straightforward to see that solutions of \ae-theory for which the aether is hypersurface-orthogonal will be solutions of the low-energy limit of Ho\v rava gravity
as well, but the converse is not necessarily true.
We also note that in Ref.~\cite{Jacobson:2010mx} it was shown that
in the low-energy limit of Ho\v rava gravity, equation~\eref{hleq}
is actually implied by equation~\eref{hl1} thanks to the Bianchi identity.
This is something that we will use in section~\ref{slowly_rotating}.

\subsection{Observational constraints and viability}

The parameters $c_i$ define a 4-dimensional parameter space for \ae-theory. Similarly, $\lambda$, $\xi$ and $\eta$ define a 3-dimensional parameter space 
for low-energy Ho\v rava gravity. In this paper, we will only focus on the regions of these parameter spaces where the theories are dynamically well-behaved and observationally viable.

\begin{figure}
\includegraphics[type=pdf,ext=.pdf,read=.pdf,width=6cm]{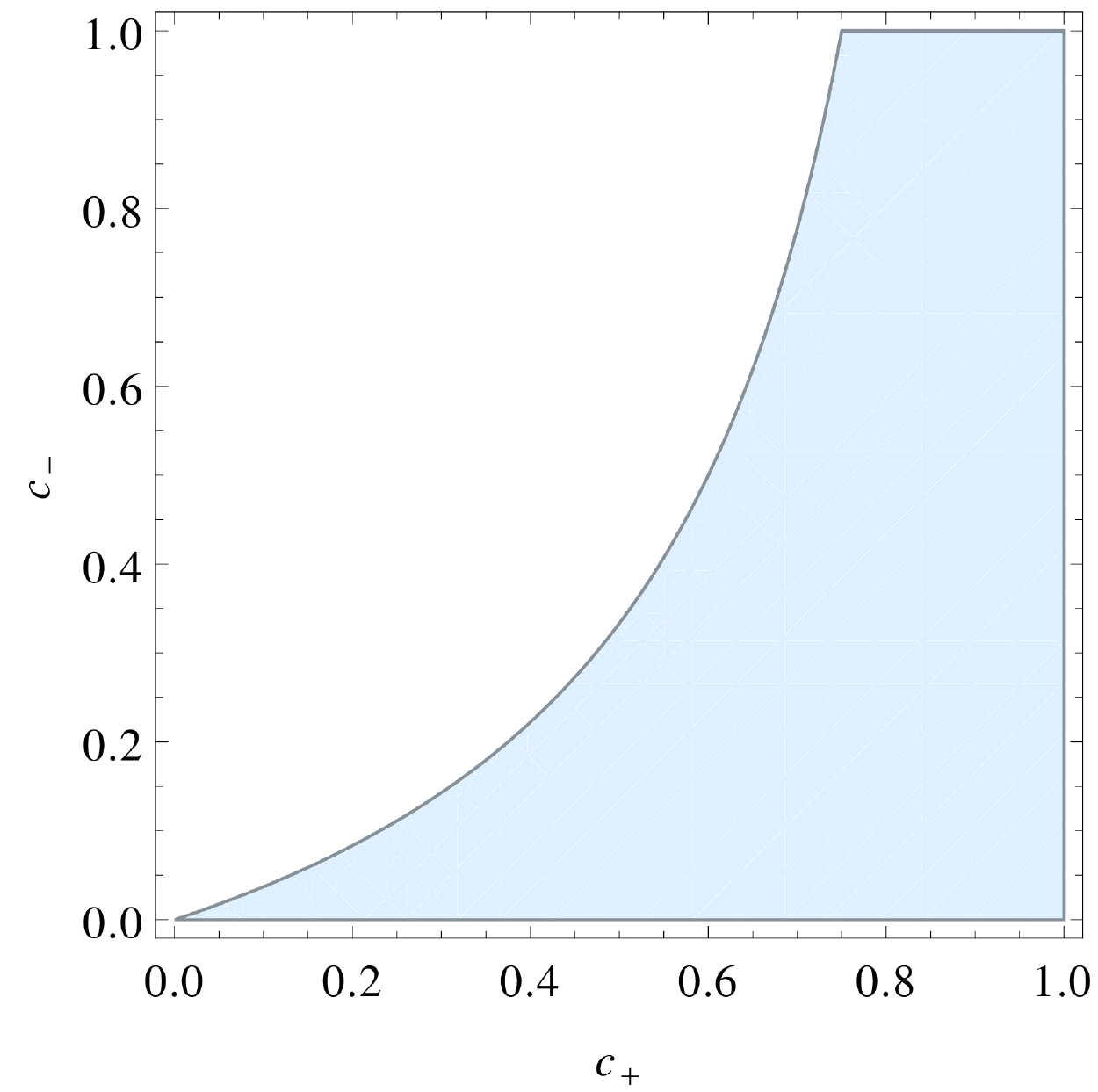}
\hskip1cm
\includegraphics[type=pdf,ext=.pdf,read=.pdf,width=6cm]{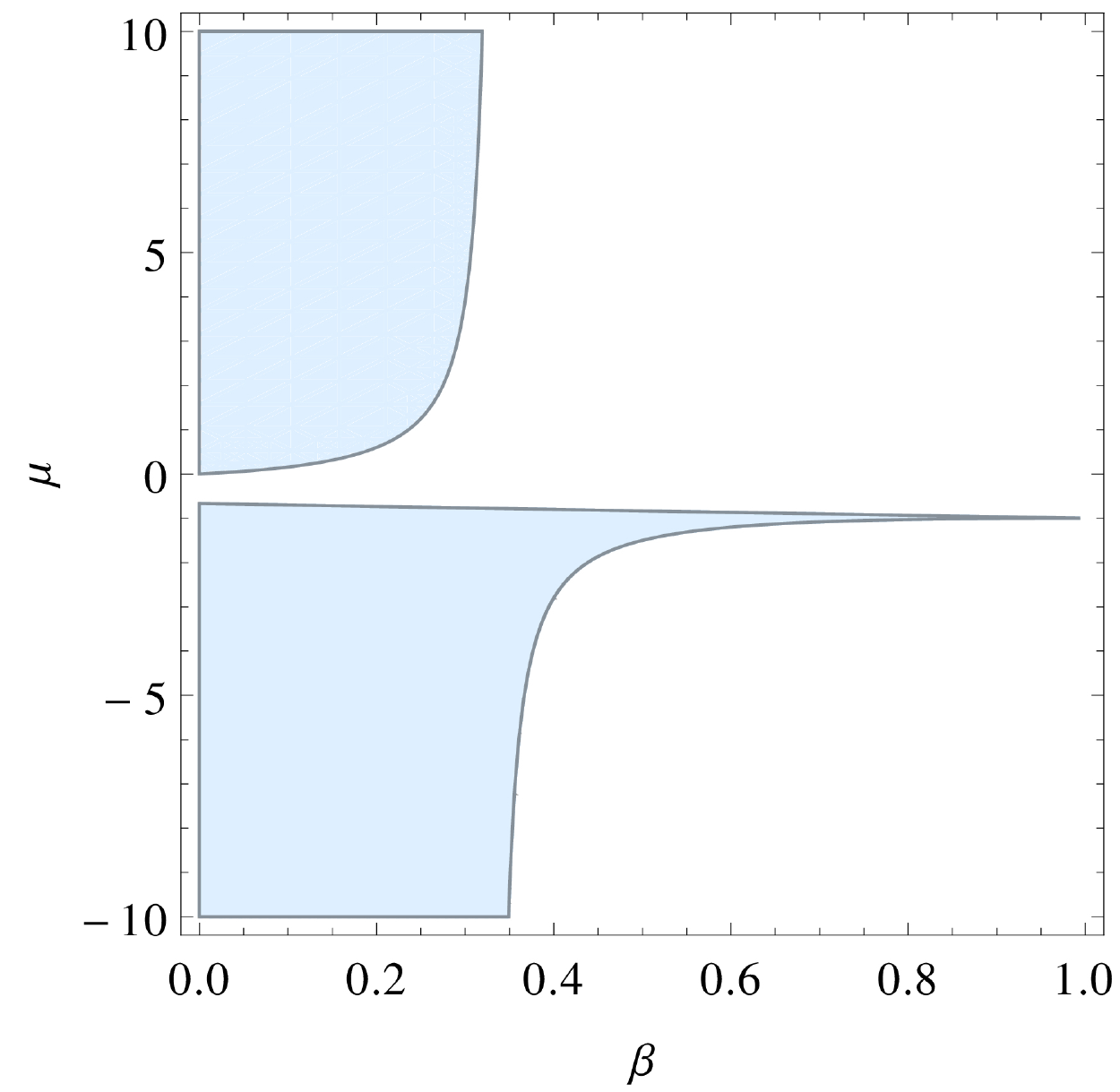}
\caption{\label{parspace} The viable regions (in light blue) of the parameter space of \ae-theory (left) and Ho\v rava gravity (right). General Relativity corresponds to the origin in both cases. For \ae-theory the viable region extends upwards in the vertical direction, and for Ho\v rava gravity both upwards and downwards in the vertical direction. }
 \end{figure}

Particularly strong constraints follow from the requirement that the two theories reproduce the weak-field predictions of General Relativity
in the presence of matter (i.e. when one adds matter fields  minimally coupled to the metric). One 
can indeed perform a post-Newtonian expansion, as in General Relativity, and cast the results
into the usual parametrized post-Newtonian (PPN) framework~\cite{Will:2005va}.
In both theories, all PPN parameters turn out to be identical to those of General Relativity except two~\cite{Blas:2010hb,Eling:2003rd,Blas:2011zd}, $\alpha_1$ and $\alpha_2$. 
These parameters measure preferred-frame effects,
and are experimentally constrained to very small values, $|\alpha_1|\lesssim 10^{-4}$ and $|\alpha_2|\lesssim 10^{-7}$,
in agreement with the predictions of General Relativity, where $\alpha_1=\alpha_2=0$. Requiring then that  $\alpha_1$ and $\alpha_2$
vanish exactly, we obtain
\be
c_2=\frac{-2 c_1^2-c_1 c_3+c_3^2}{3 c_1}\,,\quad c_4=-\frac{c_3^2}{c_1}\,,
\ee
for \ae-theory and 
\be
\eta=2\xi-2
\ee
for Ho\v rava gravity, and we are therefore left with a two-dimensional parameter space for both theories. Requiring vanishing preferred frame parameters might seem a strong requirement. However, since $\alpha_1$ and $\alpha_2$ are dimensionless, if one allows  $\alpha_1$ and $\alpha_2$ to take values that are nonzero but  satisfy  the observational constraints, then  the predictions of both theories will be close to those obtained
for $\alpha_1=\alpha_2=0$ to within an accuracy $\lesssim 10^{-4}$, and it is therefore convenient to decrease the dimension of the parameter space by
setting $\alpha_1$ and $\alpha_2$ exactly to zero.

It is sensible to ask that all of the propagating modes of the two theories have positive energy and be free of gradient instabilities in flat space. 
Additionally, there should be no modes propagating subluminally. If this were the case, matter could decay into these modes via a \v Cerenkov-like process, 
and this would lead to observable effects \cite{Elliott:2005va}. These requirements impose further constraints onto the parameters of the two theories, 
identifying only specific regions of the aforementioned 2-dimensional surface as viable. Details can be found in Ref.~\cite{Barausse:2011pu}. Here we limit ourselves to show a plot 
of the allowed part of the parameter space in Figure~\ref{parspace}. The following redefinitions of parameters allow for a better visualisation:
\be
c_\pm=c_1\pm c_3\,; \qquad \mu=\lambda/\xi-1\,, \quad \beta=(\xi-1)/\xi\,,
\ee
We stress that there are further constraints on Lorentz-violating gravity theories that could further restrict the viable parameter space.
Binary-pulsar observations have long been known to carry the potential for constraining the couplings to regions near the origin in the $(c_{+},c_{-})$ and $(\beta,\mu)$ planes~\cite{Foster:2007gr,Foster:2006az,Blas:2011zd}, but calculations have become sufficiently accurate for imposing precise constraints only very recently~\cite{sensitivities_AE}.
Similarly, cosmological observations of the cosmic microwave background and the matter power spectrum constrain the couplings to values of order unity or less~\cite{Zuntz:2008zz,Audren:2013dwa}.
Therefore, we focus here on the regions of the parameter spaces where $c_-<1$  and $|\mu|<10$, since larger values of the couplings are unlikely to be compatible
with these observations.

\section{Spherically symmetric black holes}
\label{sphBH}

In this section, we briefly review the derivation and main physical properties of spherically symmetric, static and asymptotically flat black holes in 
both low-energy Ho\v rava gravity (referred to simply as Ho\v rava gravity from now on) and \ae-theory. For a more detailed discussion, see Ref.~\cite{Barausse:2011pu}, where
these solutions were derived.

\subsection{Setup and asymptotics}

\label{setup}

First, let us note that  spherically symmetric, static and asymptotically flat  black holes are actually exactly the same in the two theories~\cite{Jacobson:2010mx,letter,Blas:2010hb,Blas:2011ni}. 
This fact is not obvious. A spherically symmetric, static aether vector field is always hypersurface-orthogonal and therefore, as discussed at the end of section \ref{aeandhl}, 
such solutions of \ae-theory will also be solutions of Ho\v rava gravity, but the converse is not generically true. 
However, it can be shown to hold for spherically symmetric, static and asymptotically flat solutions. Inserting the spherically symmetric and static ans\"atze 
$ds^2 =g_{tt}(r) dt^2+g_{rr}(r) dr^2+r^2 d\Omega^2$ and $\boldsymbol{u}= u_t(r) dt+u_r(r) dr$ (with $\boldsymbol{u}$ satisfying
the unit constraint $\boldsymbol{u}\cdot \boldsymbol{u}=1$) into equation~\eref{AEeq}, one straightforwardly obtains that $\AE^\theta=\AE^\varphi=0$ identically, while $\AE^r=0$ implies $\AE^t=0$. 
So, in order to show that any solution of Ho\v rava gravity is also a solution of \ae-theory, it is sufficient to prove that equation~\eref{hleq} implies $\AE^r=0$. This can be seen
by integrating equation~\eref{hleq} (which contains only $r$ and $\theta$ derivatives because of the spherically symmetric static ansatz)
between two arbitrary radii $r=r_1$ and $r=r_2$. Gauss's divergence theorem then gives
 \begin{equation}
\int_{\theta=0}^{\theta=\pi}\frac{1}{\sqrt{\nabla^\alpha T \nabla_\alpha T}} \sqrt{-g} \AE^r d\theta\Bigg\vert^{r=r_2}_{r=r_1}=0\,.
\end{equation}
One can then note that asymptotic flatness requires $\sqrt{\nabla^\alpha T \nabla_\alpha T}\sim 1$,
$\sqrt{-g}\sim r^2\cos\theta$ and $\AE^r\sim \partial^2 u \sim 1/r^3$, and thus
  $\sqrt{-g} \AE^r/\sqrt{\nabla^\alpha T \nabla_\alpha T}\sim 1/r \to 0$,
so sending $r_2\to\infty$ one obtains $\AE^r=0$ at the (arbitrary) radius $r=r_1$.

Therefore,
the static, spherically symmetric and asymptotically flat black-hole solutions are
indeed the same in the two theories, and this allows us to focus on \ae-theory only (because the solutions that we will find will also be all of the solutions of Ho\v rava gravity). 
In order to actually find these black-hole solutions, it is convenient to use ingoing Eddington-Finkelstein coordinates, which are regular at the horizon. In these
coordinates, the line element and the aether field take the form
\bea
\label{efmetric}
ds^2=f(r)dv^2-2 B(r)dv dr-r^2d\Omega^2\,,\\
\label{efaether}\boldsymbol{u}=A(r) \partial_v-\frac{1-f(r) A^2(r)}{2 B(r) A(r)}\,\partial_r\,.
\eea
Imposing asymptotic flatness
$f$, $B$ and $A$ have the following behavior~\cite{Barausse:2011pu,Eling:2006ec}
\begin{eqnarray}
f(r) &=& \left(1+\frac{F_1}{r}+\frac{c_{1}+c_{4} }{48} \frac{F_1^3}{r^3}+ \cdots \right) \chi^2 \label{asyF}\\
B(r) &=& \left(1+\frac{c_{1}+c_{4}}{16}  \frac{F_1^2}{r^2}-\frac{c_{1}+c_{4}}{12} \frac{F_1^3}{r^3}+\cdots \right) \chi\label{asyB} \\
A(r) &=& \left[1-\frac12 \frac{F_1}{r}+\frac{A_2}{r^2}+
\left(\frac{1}{16} F_1^3-\frac{c_1+c_4}{96}
F_1^3-F_1 A_2\right) \frac{1}{r^3}+\cdots \right]\frac{1}{\chi}\nn\\&&\label{asyA}
\end{eqnarray}
where $F_1$ and $A_2$ are  constants specifying the solution, and $\chi$ is a gauge parameter that can be set to 1 by rescaling the $v$ coordinate.
 $F_1$ is  a measure of the ``characteristic  size'' of the
solution as measured at infinity, i.e.~one can define a ``gravitational radius'' $r_g=-F_1$. This is in turn proportional to the total mass $M_{\rm tot}$ 
measured by a distant observer via $r_g=2 G_N M_{\rm tot}$, where
$G_N$ is the gravitational constant measured by Newtonian experiments (related to the ``bare'' gravitational constant $G_{\ae}$ appearing
in the action by $G_N=G_{\ae}/[1-(c_1+c_4)/2]$~\cite{Gn_vs_Gbare}). 
$A_2$ is instead an ``aether charge'', and in principle may take arbitrary values. As discussed in the next section, however,
reasonable regularity requirements force $A_2$ to take a specific value for any given $F_1$.

\subsection{Numerical implementation}
\label{numerics}

With the ans\"atze of eqs.~\eref{efmetric} and \eref{efaether} the only non-trivial  \ae-theory field equations are 
$E^{vv}=E^{vr}=E^{rr}=E^{\theta \theta}=\AE^v=0$
(note that $\AE^r$ is proportional to $\AE^v$
so it does not need to be imposed separately), so the system is equivalent to
\bea\label{eqns2}
E^{vv}=E^{\theta \theta}=\AE^v=0,\quad C^v=C^r=0\,,
\\
\label{constraint}
C^\nu\equiv 2 {E^{r\n}}+u^r\AE^\n=0\,.
\eea
The first three equations of this system can
be rearranged into a system of ``evolution'' ordinary differential
equations in the radial coordinate~\cite{Barausse:2011pu}, with schematic form 
\bea
\label{Fpp}
f''&=& f''( A, A', B, f, f')\label{ev1}\\
\label{App}
A''&=& A''( A, A', f, f')\label{ev2}\\
\label{Bp}
B'&=&B'( A, A', B, f, f')\label{ev3}\,.
\eea
The remaining equations, $C^v=C^r=0$,
can instead be viewed as initial value constraints for the evolution system \eref{ev1} -- \eref{ev3}~\cite{Barausse:2011pu,diffeo_ted}.

The existence of a set of initial value constraints 
comes about in a similar way as in General Relativity,
where diffeomorphism invariance applied to the Einstein-Hilbert action yields the
Bianchi identity, which in turn ensures the conservation
of the energy and momentum constraints under a general-relativistic evolution.
In the case of \ae-theory and Ho\v rava, one can start from the
covariant action~\eref{S}, and show~\cite{diffeo_ted} that diffeomorphism invariance 
implies the \textit{identity}
\be\label{identity}
\nabla_\m \left(2 {E^{\m\n}}+ u^\mu\AE^\n\right)= -\AE_\m\nabla^\n u^\m.
\ee
For our static spherically symmetric system, the 
right-hand side of equation~\eref{identity} is zero if the evolution 
equations~\eref{eqns2} are imposed, so setting $\nu=i$  (with $i=r,v$) equation~\eref{identity}
 becomes
\begin{equation}
\partial_r C^i+\Gamma (E-u\AE)\mbox{-terms}=0
\end{equation}
along the evolution.
(Note that we used the fact that partial derivatives with respect to
$v$ are zero due to the static ansatz.) Therefore,
if one imposes $C^r=C^v=0$ at some ``initial'' radius $r_0$,
then  $C^r=C^v=0$ also a nearby radius $r_1=r_0+\delta r$, if  
the evolution equations are used to move from $r_0$ to $r_1$~\cite{Barausse:2011pu}.
Moreover, the very structure of equation~\eref{identity} shows
that $C^r$ and $C^v$ can only depend on $B$, $A$, $A'$, $f$ and $f'$,
i.e. they do indeed only depend on initial data for the evolution system \eref{ev1} -- \eref{ev3}~\cite{Barausse:2011pu}.
In fact, if $C^r$ and $C^v$ contained higher-order derivatives (e.g. $B'$, $f''$ or $A''$), the left-hand side of
equation~\eref{identity} would depend on e.g. $B''$, $f'''$ or $A'''$,
while the right-hand side depends only on $B$, $B'$, $A$, $A'$, $A''$, $f$, $f'$ and $f''$.

Despite the presence of these two initial value constraints, the evolution system \eref{ev1} -- \eref{ev3} still
allows to freely specify three of the five initial data $A(r_0)$, $A'(r_0)$, $B(r_0)$, $f(r_0)$, $f'(r_0)$, 
where the initial radius $r_0$ can be chosen to be the metric horizon's radius $r_{\rm H}$, which satisfies
$f(r_{\rm H})=0$ but whose numerical value needs to be specified.
(Note that $f(r_{\rm H})=0$ at the horizon does not decrease
the number of initial conditions, because $C^r\propto f$.)
Therefore, generic solutions will depend on four parameters, while 
the asymptotically flat solution~\eref{asyF} -- \eref{asyA} depends on three parameters 
$F_1$, $A_2$ and $\chi$. (Obviously, one may 
set $r_{\rm H}=1$ with a choice of units, but that would also
fix the lengthscale $F_1$ in the asymptotic solution. Similarly, $\chi$ may be set to 1 by rescaling $v$, but this would
impose one additional condition on the initial data of the evolution
equations.)
 This shows that the requirement of asymptotic flatness
constrains the black-hole solutions and decreases the number of free parameters from four to three,
unlike in General Relativity, where asymptotic flatness follows from the field equations. In practice,
one just needs to impose one condition, e.g. that
$f(r)A(r)^2\to 1$ as $r\to\infty$
(c.f. equations \eref{asyF} -- \eref{asyA}),
and the field equations then ensure that the asymptotic behavior of
the solution is given by equations \eref{asyF} -- \eref{asyA}~\cite{Barausse:2011pu,Eling:2006ec}.
It would therefore seem that asymptotically flat, spherically symmetric and static black holes are characterized by three parameters. However, as
already mentioned, the gauge parameter $\chi$ in equations~\eref{asyF} -- \eref{asyA} 
can be set to 1 without loss of generality, so these black holes are actually characterized
by the length-scale $F_1$, 
or equivalently the total mass $M_{\rm tot}$, and an aether charge $A_2$.

However, this 2-parameter family of solutions has an unpleasant characteristic: for a given mass $M_{\rm tot}$ the solution 
has a finite-area singularity on the spin-0 horizon, unless $A_2$ takes a specific value ${A}^{\rm reg}_2$ \cite{Eling:2006ec,singularBHs}. 
Solutions that are singular on the spin-0 horizon are not expected to form from gravitational collapse or any other physical process,
because there is nothing special about the spin-0 horizon's location. Indeed, 
numerical work simulating gravitational collapse in spherical symmetry in
\ae-theory supports this expectation, as no singularity at the spin-0 horizon was 
found to arise~\cite{collapse_ae}. Therefore, it seems natural to impose that the solution be regular everywhere apart from the central singularity. This will select $A_2={A}^{\rm reg}_2$ and the corresponding solution will only be characterized by one parameter, $M_{\rm tot}$, as in General Relativity.

To impose this condition in practice, it is convenient to work with the spin-0 metric 
$g^{(0)}_{\a\b}=g_{\a\b}+(s_0^2-1)u_\a u_\b$ (c.f. section \ref{aeandhl}), write it 
in ingoing Eddington-Finkelstein coordinates so as to put it in the form~\eref{efmetric}, and 
follow the procedure outlined above to split the field
equations in evolution and constraint equations. One can then
solve the evolutions equations perturbatively near the spin-0 horizon's radius $r^{(0)}_{\rm H}$
(which we set to 1 with a choice of units),
and that calculation shows indeed that generic black-hole solutions are singular at $r^{(0)}_{\rm H}$, unless
a specific combination of the initial conditions vanishes~\cite{Barausse:2011pu}, thus leaving us with two initial data
for the evolution system \eref{ev1} -- \eref{ev3}, before
asymptotic flatness is imposed.\footnote{Again, the fact that
$f^{(0)}(r^{(0)}_{\rm H})=0$ does not decrease the number of initial data, as 
this condition is degenerate with $C^r=0$.}

In practice, we then eliminate one more initial condition by rescaling the time coordinate $v$ so as
to have $B^{(0)}(r^{(0)}_{\rm H})=1$, thus leaving just one free initial condition, e.g. $A^{(0)}(r^{(0)}_{\rm H})$, which
we fix by imposing asymptotic flatness (i.e. $\lim_{r\to\infty} f(r) A(r)^2=1$)
with a shooting method.\footnote{Note that $f(r) A(r)^2$ is dimensionless and invariant under a rescaling of $v$,
so it asymptotic value is not affected by our choice of units or by our choice $B^{(0)}(r^{(0)}_{\rm H})=1$.}
The parameters appearing in the asymptotic solution
\eref{asyF}--\eref{asyA}, including the aether charge ${A}^{\rm reg}_2$, can then be extracted from the numerical solution, and one is left with
a single black-hole solution in units where $r^{(0)}_{\rm H}=1$, or a one-parameter family of solutions (parametrized by the black-hole mass $M_{\rm tot}$)
in generic units~\cite{Barausse:2011pu}. While this is similar to the general-relativistic case, 
we stress that
Birkhoff's theorem is not valid in
Lorentz-violating gravity, i.e.~the metric outside a spherically symmetric static star differs from that of a spherically symmetric static black hole. In particular,
the aether has a radial component in the latter case (cf. equation~\eref{efaether}), which is absent in the former~\cite{NS_ae}.

\subsection{Solutions and astrophysical implications}

Here we will discuss the spherically symmetric static black hole solutions of Ref.~\cite{Barausse:2011pu}, whose derivation
was briefly reviewed in sections~\ref{setup} and \ref{numerics}. We focus on the spacetime outside the metric event horizon. While that is clearly
the most important region for astrophysical purposes, the presence of instantaneous modes in the UV limit of Lorentz-violating gravity theories may in principle 
allow  sufficiently high-energy phenomena to probe 
regions inside the event horizon  (c.f. section~\ref{intro}).
This possibility will be discussed in section~\ref{Uhor}.

As far as the geometry outside the metric horizon is concerned, two quantities are of particular relevance for observations of astrophysical black holes.
The first is the location of the innermost stable circular orbit (ISCO), which determines the inner edge of thin accretion disks 
and their radiative efficiency~\cite{thindisks}. The location of the ISCO, more specifically, affects the measurements of both the continuum X-ray spectrum 
and the fluorescent iron K-$\alpha$ emission lines of the accretion disk (see e.g. Refs.~\cite{continuum,iron}), which are routinely used to estimate the spin of black-hole candidates. 
More recently, it has been
shown that such measurements can in principle allow the very geometry of  black-hole candidates to be tested, {\it i.e.}~by highlighting possible deviations 
away from the Kerr metric predicted by General Relativity~\cite{psaltis_iron,BB10,BB11}. 
Although these tests of the geometry of black-hole canditates are still not feasible with current data (which is affected by systematic errors hard to quantify),
they may become possible with future X-ray detectors. It is therefore interesting to understand, at least in the simple spherically symmetric, static case, how
much the radius of the ISCO of black holes (and, hence, accretion disk spectra) change in Lorentz-violating gravity.

The second quantity of astrophysical interest is the impact parameter of the photon circular orbit, which measures the cross section of photons
on the black hole, and which therefore regulates the ``shadow'' of black holes (illuminated e.g. by their accretion disk), or more in general any light-deflection
experiments in the strong-field regions of black holes. Just as in the case of X-ray spectra, such experiments can in principle be used to test the Kerr nature of black-hole 
candidates, and especially in the case of SgrA${}^*$ (our Galaxy's massive black hole)
this may become possible in the near future thanks to very-long baseline interferometric observations~\cite{shadows1,shadows2}. Hence, 
it is important to assess the magnitude of the expected deviations away from General Relativity in Lorentz-violating gravity theories.

\begin{figure}
\includegraphics[type=pdf,ext=.pdf,read=.pdf,width=6cm]{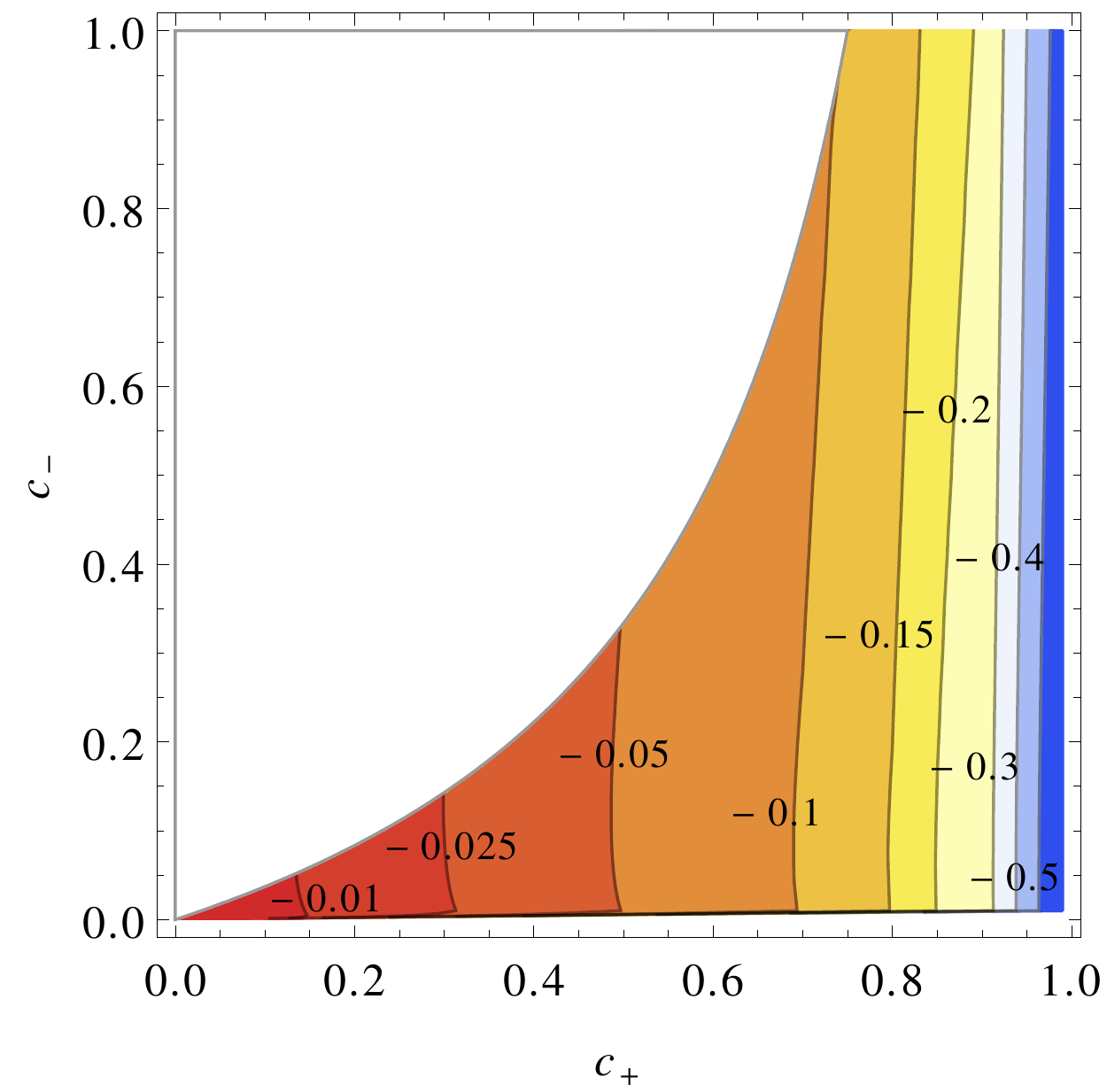}
\hskip 1cm
\includegraphics[type=pdf,ext=.pdf,read=.pdf,width=6cm]{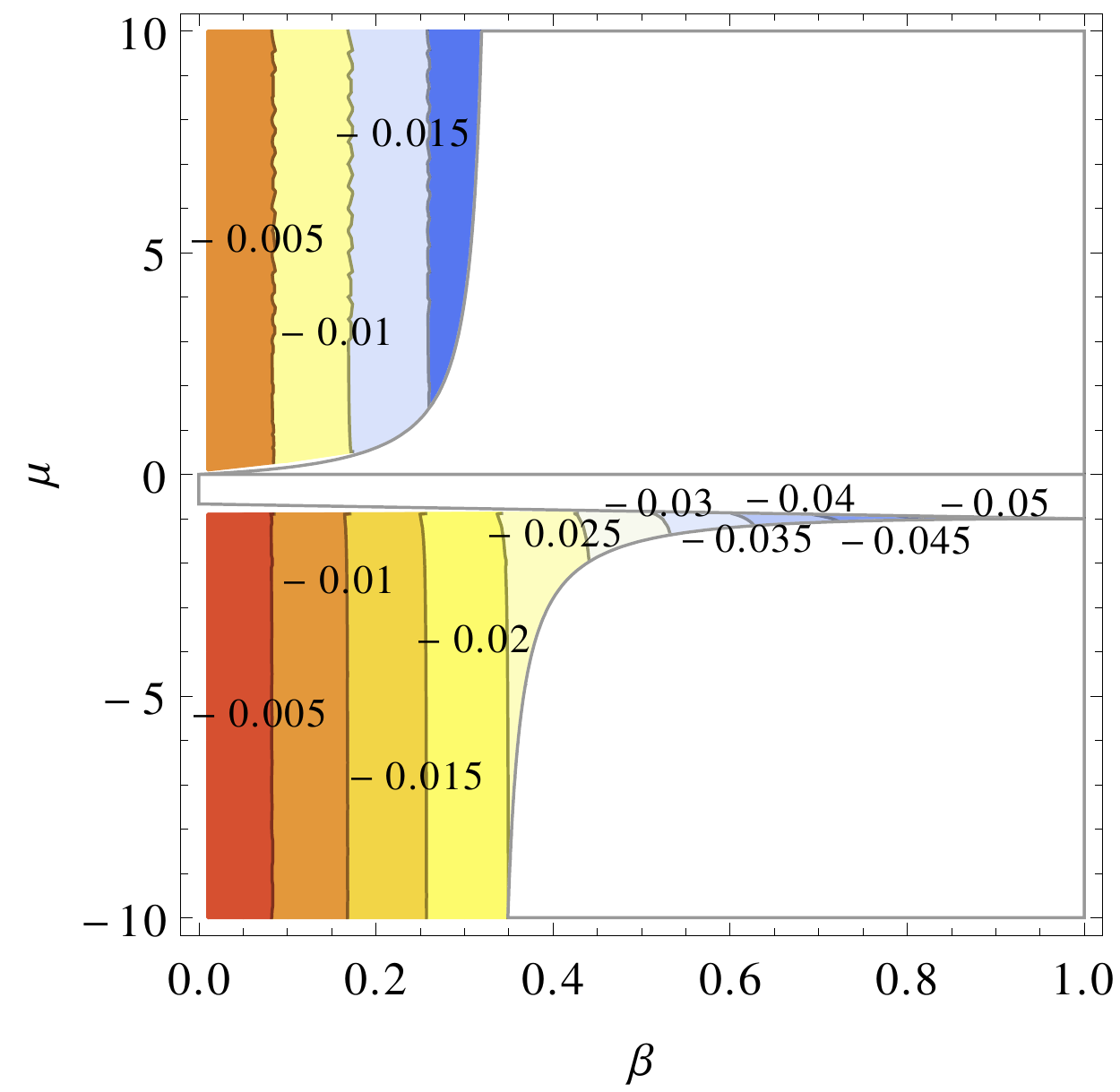}
\caption{\label{isco} 
The fractional deviation away from General Relativity for the dimensionless product $\omega_{_{\rm ISCO}} r_g$, for 
\ae-theory (left) and low-energy Ho\v rava gravity (right), in the viable region of the parameter plane. Negative values 
mean that $\omega_{_{\rm ISCO}} r_g$ is smaller in Lorentz-violating gravity than in General Relativity.
These figures are re-rendered versions of Figures 2 and 6 of Ref.~\cite{Barausse:2011pu}.}
\end{figure}

\begin{figure}[h]
\includegraphics[type=pdf,ext=.pdf,read=.pdf,width=6cm]{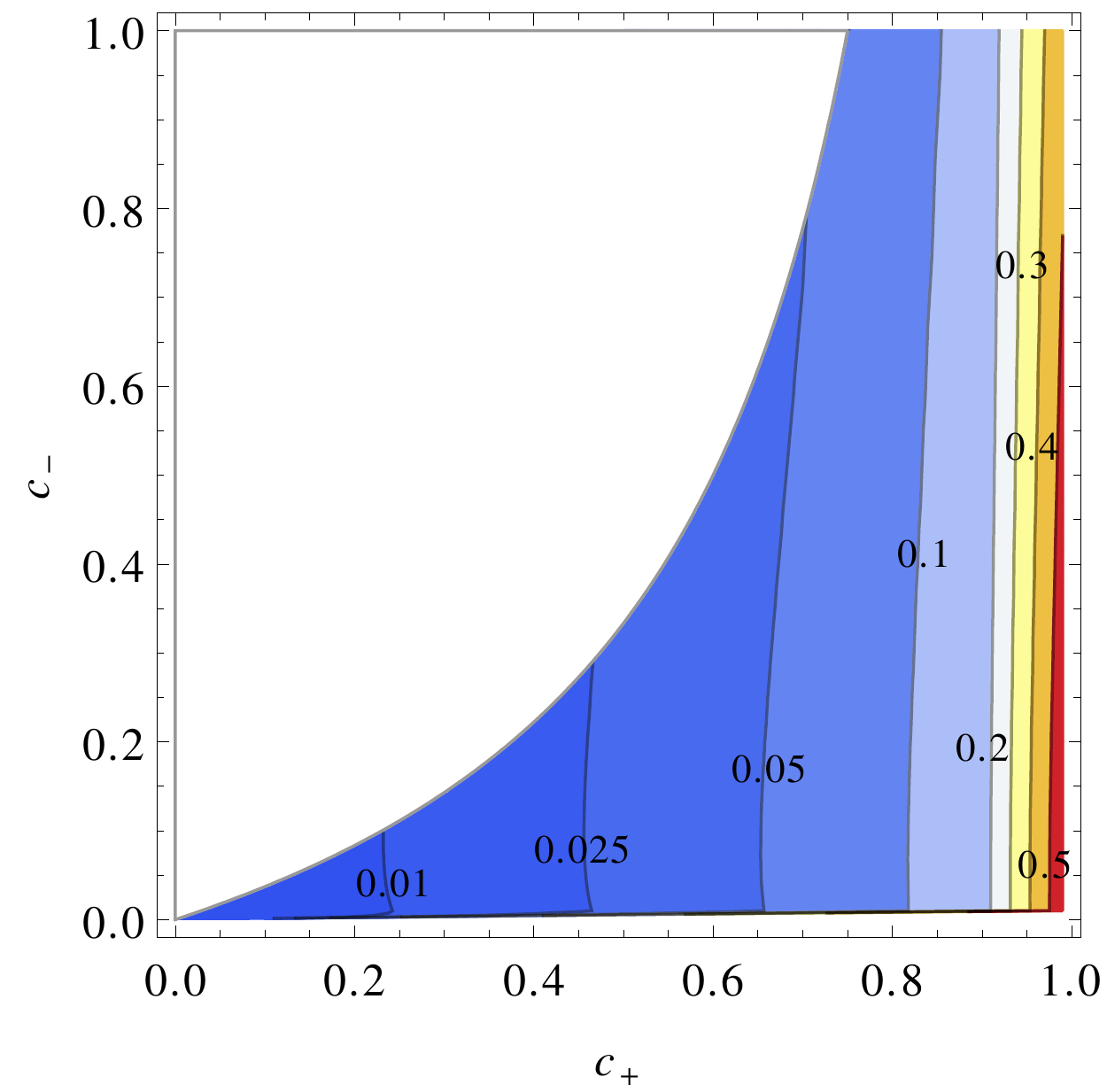}
\hskip 1cm
\includegraphics[type=pdf,ext=.pdf,read=.pdf,width=6cm]{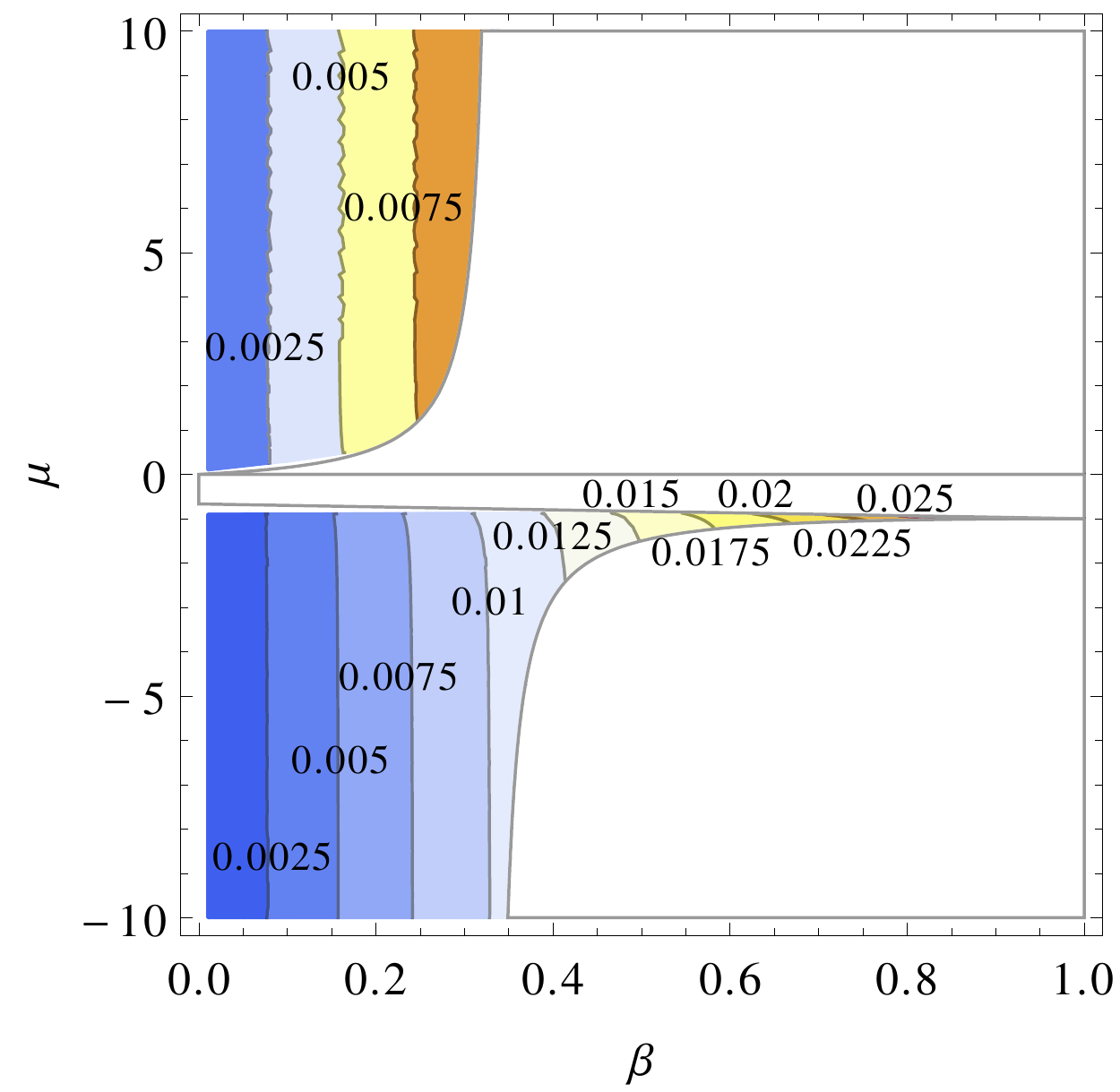}
\caption{\label{bph} The same as in Figure~\ref{isco}, but for the dimensionless quantity $b_{\rm ph}/r_g$. Positive values mean that this
quantity is larger in Lorentz-violating gravity than in General Relativity.
These figures are re-rendered versions of Figures 4 and 8 of Ref.~\cite{Barausse:2011pu}.}
\end{figure}

In order to obtain results that are independent of the coordinates used in our black-hole solutions, we look at gauge-invariant quantities, namely
\textit{(i)} the dimensionless product between the ISCO frequency $\omega_{_{\rm ISCO}}$ and the gravitational radius $r_g$; \textit{(ii)} the gauge invariant impact parameter $b_{\rm ph}$ of the photon circular orbit, defined
as $b_{\rm ph}=L_{\rm ph}/E_{\rm ph}$ (with $L_{\rm ph}$ and $E_{\rm ph}$ being the circular photon orbit's 
angular momentum and energy as measured by an asymptotic observer), normalized to the the gravitational radius $r_g$.
(Note that $b_{\rm ph}$ also measures the orbital frequency $\omega_{\rm ph}$ of the circular photon orbit, as $\omega_{\rm ph}=1/b_{\rm ph}$.)
We plot fractional deviations of both quantities aways from their general-relativistic values (which are $\omega_{_{\rm ISCO}} r_g=2\cdot6^{-3/2}$ and $b_{\rm ph}/r_g=3 \sqrt{3}/2$)
in Figures \ref{isco} and \ref{bph}. As can be seen, the deviations are typically of a few percent, and therefore too small to be constrained
with present data~\cite{BB10}. On the other hand, if we were able to generalize our results to spinning black holes (see section~\ref{slowly_rotating} for some work in this direction),
future experiments might offer better prospects for these tests, at least in some regions of parameter space and provided that systematics are well understood.
In fact, Figures \ref{isco} and \ref{bph} show that for a given total mass, black holes tend to be ``larger'' in
Lorentz-violating gravity than in General Relativity, \textit{i.e.} both the impact parameter for photons, $b_{\rm ph}$,
and the ISCO gauge-invariant radius, $R_{\rm ISCO}=(M/\omega^2)^{1/3}$, increase relative to General Relativity. Because black-hole spins tend
to decrease $R_{\rm ISCO}$ and $b_{\rm ph}$, one might be able to break the degeneracy between the effect of the spin and that of Lorentz violations.
(Consider that the measurements of the spins of black-hole candidates typically yield quite large values~\cite{continuum,iron}, especially in the case
of supermassive black holes, so it might be challenging to explain existing observations in theories with large degrees of
Lorentz violation in the gravity sector.)  

To detect such small deviations away from General Relativity, however, the ideal tool would be a space-based gravitational-wave interferometer such as the European project eLISA~\cite{elisa},
which would be able to detect deviations away from the Kerr metric down to fractional differences of $\sim 10^{-6}$. This would be achieved by detecting gravitational radiation
from extreme mass-ratio inspirals (EMRIs), i.e. systems comprised of a stellar-mass black hole or a neutron star, orbiting around a massive black hole with mass $10^5-10^6 M_\odot$. To perform
these tests, however, a thorough study of gravitational-wave emission in Lorentz-violating gravity is needed~\cite{sensitivities_AE}.

\subsection{Interior and universal horizons}

\label{Uhor}

\begin{figure}[t]
\includegraphics[type=pdf,ext=.pdf,read=.pdf,width=5.2cm]{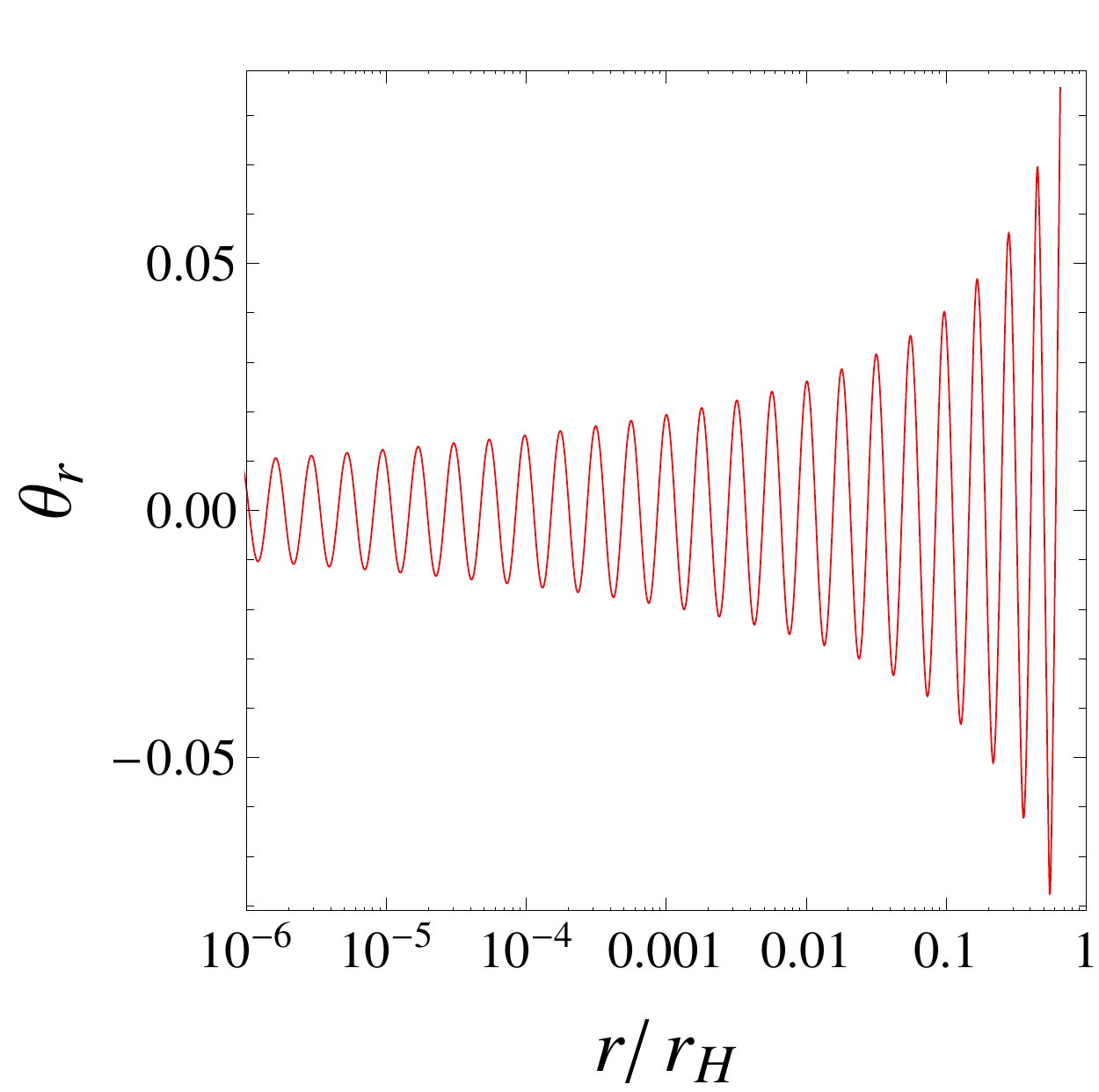}\;\;\;\;
\includegraphics[type=pdf,ext=.pdf,read=.pdf,width=7.43cm]{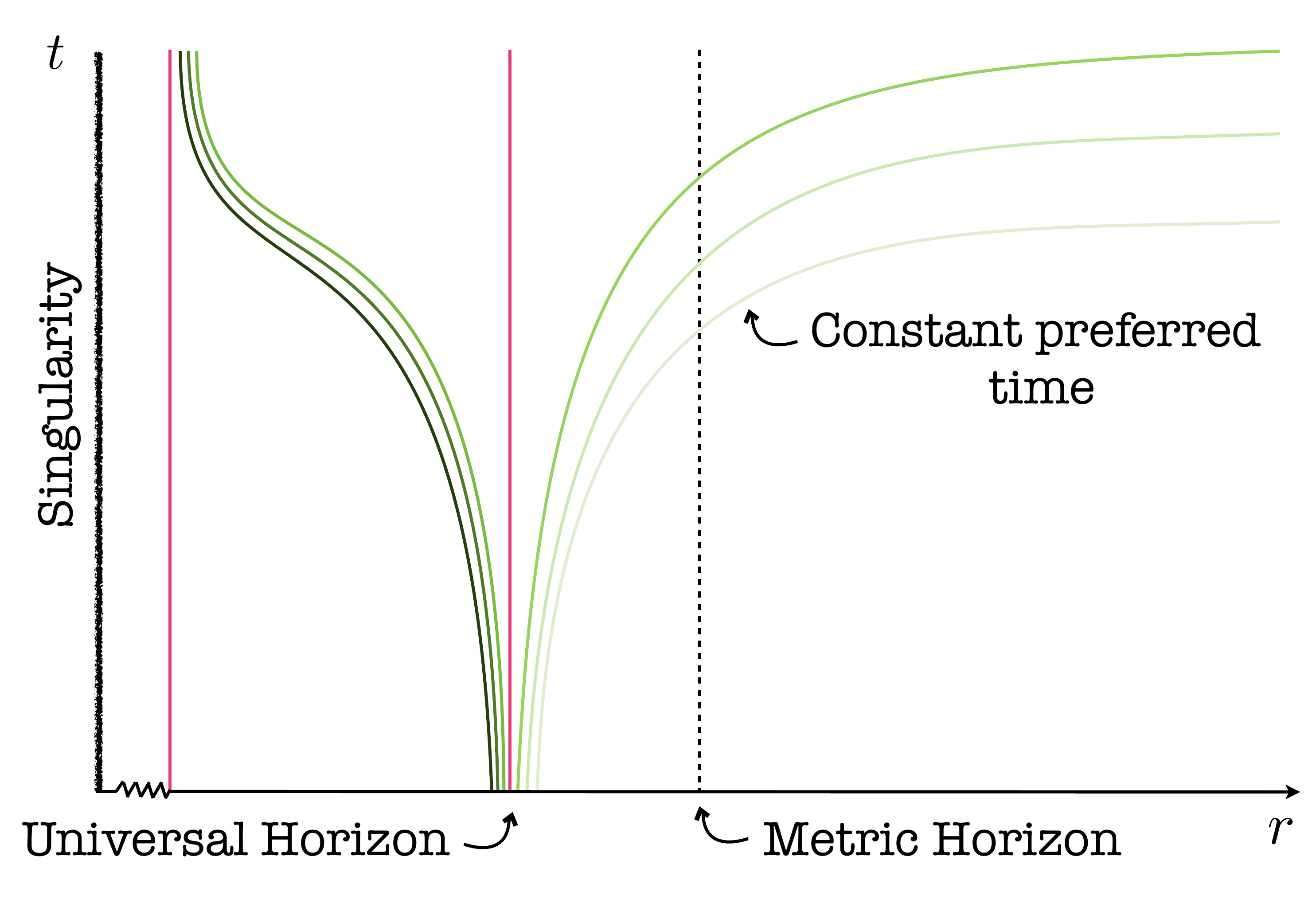}
\caption{\label{angle} {\it Left panel:} the boost angle between the aether and the future-directed normal to the $r=$ constant hypersurfaces 
for \ae-theory with $c_+=0.8$, $c_-=0.01$. $\theta_r$ vanishes for the first time soon after the metric horizon (located at $r=r_H$) is crossed, 
and the aether becomes normal to an $r=$constant hypersurface pointing inwards. 
{\it Right panel:} schematic space-time diagram of a black hole with a universal horizon. The green curves represent constant preferred-time hypersurfaces, 
and they get darker as one moves into the future. These curves span the exterior of the universal horizon (red line), but none of them crosses it. 
The curves that span the region inside the universal horizon do not extend into the exterior. 
This implies that any signal emitted in the interior will not be able to escape to the exterior, simply by the requirement that it travel into the future. 
More universal horizons may be present at smaller radii but we have truncated the region of the diagram between the second one and the singularity.
}
\end{figure}

Outside the metric horizon, the black holes that we have studied in the previous section are very similar to the Schwarzschild solution of
General Relativity, as can be seen from Figures~\ref{isco} and \ref{bph}. We will now turn our attention to the interior regions
of those solutions. As expected, the geometry is singular at the center, as can be verified by looking at the behavior of the 
Kretschmann scalar as $r\to0$~\cite{Barausse:2011pu}. 
It is more interesting, however, to explore the behavior of the aether in the interior, and specifically its orientation. 
To this purpose, one can calculate the boost angle between the aether and the (future-directed) normal to the constant-$r$ hypersurfaces (which are spacelike inside the metric horizon),
\be
\theta_r={\rm arccosh} {\left({\bf u} \cdot  \frac{{\rm d} r}{\sqrt{g^{rr}}}\right)}\,.
\ee
The crucial fact here is that the boost angle generically vanishes on an $r=$ constant hypersurface very close to the metric horizon (and many times again further inside)~\cite{Barausse:2011pu}. 
We show a representative example in the left panel of Figure~\ref{angle}.

Let us explain the significance of this fact. The aether is hypersurface orthogonal in our solutions (in 
Ho\v rava gravity as a generic feature of the theory, and in \ae-theory as a consequence of the static, spherically symmetric ansatz). 
A hypersurface-orthogonal aether field defines a preferred foliation by 
singling out hypersurfaces orthogonal to it. These can be thought of as hypersurfaces of constant preferred time $T$, as defined in equation~\eref{ho}. 
The aether's orientation also determines the future time direction unambiguously in this foliation. Hypersurfaces of constant $r$ have topology $S^2\times R$. 
On the other hand, $T=$ constant hypersurfaces define the foliation and, as such, they do not intersect. This implies that if a
hypersurface of constant $T=T_u$ is also one of constant $r=r_u$, 
then any hypersurface of constant $T>T_u$ (i.e. 
any hypersurface that lies in the future of $T=T_u$)
cannot extend to $r\geq r_u$. This is explained graphically in 
the right panel of Figure~\ref{angle}. We therefore 
call the outermost hypersurface for which $\theta_r=0$ {\it universal horizon}, 
as any perturbation generated inside it cannot reach the exterior, simply because it has to move into the future \cite{sergey}. 
 
This demonstrates that no signal (or information) can escape the interior of the universal horizon and propagate to the exterior. 
Remarkably, this is true irrespective of the propagation speed. 
In particular, this new type of horizon acts as a causal boundary even for modes that satisfy higher-order dispersion relations (c.f.~equation~\eref{mdsr}), 
and which can penetrate usual horizons thanks to their propagating speed diverging in the UV. 
The existence of a universal horizon is a strong indication that the notion of a black hole still makes sense 
in Lorentz-violating theories. 

Indeed, let us return to the full version of Ho\v rava gravity, and take into account the $L_4$ and $L_6$ terms in the action \eref{SBPSHfull}, 
as it is the presence of these terms that leads to higher-order dispersion relations. The universal horizon in our solutions lies close to the metric horizon (c.f. left panel of Figure~\ref{angle}),
{\it i.e.}~in a region of relatively low curvature $\sim 1/(G_{\ae} M)^2$ (for sufficiently high black-hole mass $M$). Therefore, we
do not expect the  $L_4$ and $L_6$ terms to produce any significant corrections to that part of the spacetime.
In fact, the presence of those terms will  introduce corrections 
on the order of $1/(G_{\ae} M M_\star)^2\sim M_{\rm Planck}^4/(M M_{\star})^2$. Taking $M_\star\sim 10^{-3}$ eV, which is the lowest value allowed by observations (as discussed earlier), would yield correction of order   $10^{-14} (M_{\odot}/M)^2$,
and thus negligible for astrophysical black holes.

Similar results regarding the existence of a universal horizon have also been derived in Ref.~\cite{Blas:2011ni} in the small coupling limit, which effectively also leads to decoupling and vanishing back-reaction from the aether. That is, the author explored analytically the configuration of the aether around a Schwarzschild black hole. 
The results reported in Ref.~\cite{Blas:2011ni} are in full agreement with the results of Ref.~\cite{Barausse:2011pu}, which we reviewed here. 
Aether perturbations around solutions with a universal horizon were also considered in Ref.~\cite{Blas:2011ni} (again in the decoupling limit). 
At the linear level, the solutions appear to be stable. Remarkably, however, in the case of Ho\v rava gravity, clues of possible nonlinear instabilities have been reported:
linear perturbations contain a mode that would be non-analytic on the universal horizon, but 
has no sources (at linear order) and is therefore forced to vanish because of the boundary conditions. 
However, non-linearities can actually source linear perturbations, in which case the presence of this mode would lead to a non-linear instability.

A full nonlinear analysis, or an analysis that goes beyond the decoupling limit (and, hence, takes into account the back-reaction of the aether on the metric) has
not yet been performed. It would also be interesting to understand whether the $L_4$ and $L_6$ terms can crucially affect the behavior of the perturbations. 
Clearly, the stability of the universal horizon in Ho\v rava gravity, as well as the fate of black holes if the universal horizon is indeed unstable, are important open questions. 
For instance, while spherical gravitational collapse has been shown to lead to solutions with universal horizons~\cite{collapse_ae}, 
it is unclear what would happen for a non-spherical collapse. Also, the existence and stability of the universal horizon 
is central for the thermodynamics of black holes in Lorentz-violating theories \cite{Blas:2011ni}. This topic goes beyond the scope of this volume, 
but we refer the reader to the literature for further reading \cite{Dubovsky:2006vk,Eling:2007qd,Blas:2011ni,Berglund:2012bu,Berglund:2012fk,Cropp:2013zxi}.

\section{Slowly rotating black holes}
\label{slowly_rotating}

Astrophysical black-hole candidates are known to present a variety of spins. In the case of the ``stellar-mass'' black holes (i.e.~having a mass of $5$ -- $20 M_\odot$)
believed to power {\it e.g.}~microquasars in Galactic binaries, the spin is expected to be close to its natal value (see e.g. Refs.~\cite{natal}), i.e. it probably did not change significantly 
after the formation of the black hole. (This is essentially because 
the age of its stellar companion is typically smaller
than the Salpeter timescale $\sim 5\times 10^7$ yrs on which accretion takes place at the Eddington rate.) As already mentioned, the spins of these black holes
can be measured by analyzing the X-ray spectrum of their accretion disks -- either by fitting the continuum spectrum, or by measuring the profile of the iron-K$\alpha$ fuorescent emission lines
-- and these techniques provide a variety of spin magnitudes (see e.g. Refs.~\cite{continuum,iron}).
Spin measurements for massive black-hole canditates (with $M= 10^5$--$10^9 M_{\odot}$) are instead more complicated, but iron-K$\alpha$ line profiles seem
to typically point at rather large spins (see e.g. Ref.~\cite{iron_smbh}), while comparisons between the mass function of massive black holes
and the luminosity function of quasars seems to suggest smaller but non-zero spins~\cite{soltan_argument}. Moreover, our current understanding of galaxy formation requires that massive black holes gather most
of their mass via accretion and mergers, which both produce non-zero spins~\cite{spin_evolution}, and many astrophysical models invoke black-hole spins to produce the relativistic jets
observed e.g. in Active Galactic Nuclei~\cite{BZeffect}.
Because spins are so ubiquitous in the Universe, the existence of spinning black-hole configurations is
a natural testing ground for Lorentz-violating theories. 

The most general slowly rotating, stationary, axisymmetric metric, in a suitable coordinate system, is given by~\cite{hartle_thorne}
\begin{eqnarray}\label{metric}
ds^2&=&f(r) dt^2 -\frac{B(r)^2}{f(r)}dr^2-r^2(d\theta^2+\sin^2\theta \,d\varphi^2)\nonumber\\
&&+\epsilon 
r^2 \sin^2\theta \,\Omega(r,\theta) dtd\varphi+{\cal O}(\epsilon^2)\,,
\end{eqnarray}
where $\epsilon$ is the perturbative (``slow-rotation'') parameter.
(This ansatz for the metric can be understood based on its transformation properties under $t\to-t$ and $\varphi\to-\varphi$).
Because we want to describe 
black holes, $f(r)$ and $B(r)$ are given by the spherically symmetric static solutions of Ref.~\cite{Barausse:2011pu} reviewed in the previous section.
Similarly, based again on the transformation properties  under $t\to-t$ and $\varphi\to-\varphi$ and on the unit constraint $u_\mu u^\mu=1$, we can 
write the aether 4-velocity as
\begin{eqnarray}\label{aether}\boldsymbol{u}&=& \frac{1+f(r) A(r)^2}{2 A(r)}dt
+ \frac{B(r)}{2A(r)}\left(\frac{1}{f(r)}-A(r)^2\right) dr\nn\\&&+\epsilon \frac{1+f(r) A(r)^2}{2 A(r)} \Lambda(r,\theta) \sin^2\theta d\varphi+{\cal O}(\epsilon)^2\,,
\end{eqnarray}
 where $A$ is the aether component $u^v$ in ingoing Eddington-Finkelstein coordinates (see section \ref{setup}). 
 Note that the aether's specific angular momentum is $u_\varphi/u_t=\Lambda(r,\theta) \sin^2\theta$, 
 while the aether's angular velocity is $u^\varphi/u^t\equiv \omega(r,\theta) = \Omega(r,\theta)/2-\Lambda(r,\theta) f(r)/r^2$.

In what follows, we will first focus on \ae-theory and show that if one imposes asymptotic flatness and uses the field equations, in a suitable coordinate system the 
quantities $\Omega$, $\Lambda$ and $\omega$ are actually functions of $r$ only. As a result, the field equations reduce to a system of ordinary differential equations, which
can in principle be solved numerically to obtain slowly rotating black holes in \ae-theory. We defer, however, such a study to subsequent work~\cite{ian_in_prep},
and restrict our attention here to using that set of  ordinary differential equations to show that \ae-theory slowly rotating black holes do not admit a global preferred-time slicing,
and therefore do not seem to admit a universal horizon in the sense defined in section~\ref{Uhor}~\cite{letter,arxiv_note}.
We will then move to Ho\v rava gravity, and show that the theory, unlike \ae-theory, does admit slowly rotating black holes with a universal horizon, and we 
will write down such solutions explicitly, as already derived in Ref.~\cite{arxiv_note}.

\subsection{\AE-theory}
\label{spin_ae}

Focusing, as just mentioned, on \ae-theory, let us start from the ${r\varphi}$-component of the Einstein equations \eref{EinsteinAE},
which is given by:
\begin{eqnarray}\label{eq2}
\zeta_0 \,\omega(r,\theta) &&+ \frac{({c_1}+{c_3}) \left[A(r)^4 f(r)^2-1\right]}{8 r^2 A(r)^2 B(r) f(r)} \times \nonumber\\&& \times \left[\partial^2_\theta\omega(r,\theta)
+3 \cot\theta \partial_\theta\omega(r,\theta)\right]=0\,
\end{eqnarray}
where $\zeta_0$ is a complicated expression involving the couplings $c_i$ as well as $f(r)$, $B(r)$, $A(r)$ and their derivatives.

Let us now observe that by transforming the background (i.e. the spherically symmetric static solution) under the coordinate
change $\varphi'=\varphi+\Omega_0 t/2$ [$\Omega_0={\cal O}(\epsilon)$ being a constant], we obtain $\Omega=2\omega=\Omega_0$ and $\Lambda(r,\theta) ={\cal O}(\epsilon^2)$. 
Because this coordinate-transformed spherically symmetric static solution must still be a solution of the field equation, from equation~\eref{eq2} we obtain that
$\zeta_0$ must evaluate to zero once we use the background spherically symmetric static solution. 
We have indeed verified this fact using the solutions of Ref.~\cite{Barausse:2011pu}. As a result, equation~\eref{eq2} becomes
\begin{equation}\label{eq2bis}
\frac{({c_1}+{c_3}) \left[A(r)^4 f(r)^2-1\right]}{8 r^2 A(r)^2 B(r) f(r)} \times \left[\partial^2_\theta\omega(r,\theta)
+3 \cot\theta \partial_\theta\omega(r,\theta)\right]=0\,,
\end{equation}
and the only solution to this equation that is regular at the poles is $\omega(r,\theta)=\psi(r)$.

With $\omega=\psi(r)$, the remaining non-trivial field equations are the 
the $\varphi$-component of the aether equation \eref{AEeq2}
and the ${t\varphi}$-component of the Einstein equations \eref{EinsteinAE}, 
which can respectively be written as
\begin{eqnarray}&&
\zeta_1 \Lambda(r,\theta) + \zeta_2 \psi(r)+\sum_{i=1}^4 a_i v^i\nonumber\\&&+\frac{({c_1}-{c_3}) \left[A(r)^2 f(r)+1\right]}{4 r^4 A(r)}\times  
\left[\partial^2_\theta\Lambda(r,\theta
   )+3 \cot\theta \partial_\theta\Lambda(r,\theta )-2 \Lambda(r,\theta)\right]\,,\nn\\\label{eq0}&&\\
&&\zeta_3 \Lambda(r,\theta) + \zeta_4 \psi(r)+\sum_{i=1}^4 b_i v^i\nonumber\\&&-\frac{1}{2r^4}\left[\partial^2_\theta\Lambda(r,\theta
   )+3 \cot \theta \partial_\theta\Lambda(r,\theta )-2 \Lambda(r,\theta)\right]=0\,,\label{eq1}
\end{eqnarray}
where $v=[\psi'(r), \psi''(r),\partial_r\Lambda(r,\theta),\partial_r^2\Lambda(r,\theta)]$;
 $\zeta_1$, $\zeta_2$, $\zeta_3$, $\zeta_4$ are complicated expressions involving the couplings $c_i$ as well as $f(r)$, $B(r)$, $A(r)$ and their derivatives,
and the coefficients $a_i$ and $b_i$ are functions of radius only (involving again the couplings $c_i$, as well as $f(r)$, $B(r)$, $A(r)$ and their derivatives).

Again, it is easy to show that  $\zeta_1$, $\zeta_2$, $\zeta_3$, $\zeta_4$ must evaluate to zero once the background spherically symmetric static solution is used. 
In fact, by transforming that solution under a coordinate change $\varphi'=\varphi+\Omega_0 t/2$ (with $\Omega_0$ being a constant ${\cal O}(\epsilon)$), one obtains $\Omega=2\psi=\Omega_0$
and $\Lambda={\cal O}(\epsilon)^2$. Replacing in equations~\eref{eq0} and \eref{eq1} we obtain that $ \zeta_2 $ and $ \zeta_4 $ must evaluate to zero. 
Likewise, transforming the spherical solution
under $t'=t-\ell \varphi$ [$\ell={\cal O}(\epsilon)$ being a constant],  we obtain $\psi={\cal O}(\epsilon^2)$ and $u_\varphi/u_t=\Lambda(r,\theta) \sin^2\theta =\ell$. 
Replacing this in equation~\eref{eq0} and \eref{eq1} we conclude that $ \zeta_1 $ and $ \zeta_3 $ must also evaluate to zero.\footnote{Again, we have verified explicitly 
that $\zeta_1$, $\zeta_2$, $\zeta_3$, $\zeta_4$  evaluate to zero using the solutions of Ref.~\cite{Barausse:2011pu}.}

Combining equations~\eref{eq0} and \eref{eq1} to eliminate the terms depending on the $\theta$-derivatives, one obtains
\begin{equation}\label{eqCombo}
d_1(r) \psi'(r)+d_2(r) \psi''(r)+d_3(r) \partial_r\Lambda(r,\theta)+d_4(r) \partial^2_r\Lambda(r,\theta)=0
\end{equation}
with
\begin{eqnarray}
d_1(r)&=&\frac{c_1+c_4}{4 r^5 A(r)^2 B(r)^3} \times \nonumber\\&&\Big\{r A(r)^2 B(r) f(r) A'(r)-r B(r) A'(r)+A(r)^3 \times\nonumber\\&&\Big[\left(c_1+c_3-1\right) s_1^2 f(r) \left(r B'(r)-4
   B(r)\right)+r B(r) f'(r)\Big]\nonumber\\&&-\left(c_1+c_3-1\right) s_1^2 A(r) \left[4 B(r)-r B'(r)\right]\Big\}\,,
\end{eqnarray}
where $s_1^2=({c_1^2-2 c_1-c_3^2})/[{2 \left(c_1+c_3-1\right) \left(c_1+c_4\right)}]$ is the square of the speed of the spin-1 mode,
\vskip-0.5cm
\begin{equation}
d_2(r)=-\frac{s_1^2 \left(c_1+c_3-1\right) \left(c_1+c_4\right) \left(A(r)^2 f(r)+1\right)}{4 r^4 A(r) B(r)^2}\,,\\
\end{equation}
\vskip-0.5cm
\begin{eqnarray}
&& \nn d_3(r)=\frac{1}{32 r^6 A(r)^4 B(r)^3} \left\{2 B(r)\! \left[\left(A(r)^2 f(r)-1\right)\! A'(r) \times \right.\right. \\\nonumber &&  \left(\left(\left(c_1-c_3\right)
   \left(c_1+2 c_3-c_4\right)+4 c_4\right) (1+A(r)^4 f(r)^2)\right.\\&&\nonumber \left. +2 \left(c_1^2+\left(c_3-c_4-2\right) c_1-2 c_3^2+\left(c_3+2\right) c_4\right) A(r)^2 f(r)
   \right)\\ \nonumber &&+A(r)^3
   f'(r) \left(\left(\left(c_1-c_3\right) \left(c_1+2
   c_3-c_4\right)+4 c_4\right)  A(r)^4 f(r)^2 \right.\\\nonumber && -2
   \left(c_1-c_3\right) \left(c_1+c_4\right) A(r)^2 f(r)-3 c_1^2+c_1 \left(c_3-c_4+4\right)\\
&&\nonumber \left. \left. +c_3
   \left(2 c_3+c_4\right)\right)\right]-\left(c_1^2-c_3^2+2 c_4\right) A(r) \left[A(r)^2
   f(r)-1\right]^2 \\ && \left.  \times\left[A(r)^2 f(r)+1\right] B'(r)\right \} \,,
\end{eqnarray}
%
\begin{eqnarray}
d_4(r)&=&\frac{c_1^2-c_3^2+2 c_4}{32
   r^6 A(r)^3 B(r)^2}\times\left[A(r)^2 f(r)-1\right]^2 \left[A(r)^2 f(r)+1\right]\,.
\end{eqnarray}
The system is closed by considering, besides equation~\eref{eqCombo}, either equation~\eref{eq0} or equation~\eref{eq1}. Choosing equation~\eref{eq1}, its explicit form reads
\begin{eqnarray}\label{eq1bis}
&&b_1(r) \psi'(r)+b_2(r) \psi''(r)+b_3(r) \partial_r\Lambda(r,\theta)+b_4(r) \partial^2_r\Lambda(r,\theta)\nonumber\\&&-\frac{1}{2r^4}\left[\partial^2_\theta\Lambda(r,\theta
   )+3 \cot \theta \partial_\theta\Lambda(r,\theta )-2 \Lambda(r,\theta)\right]=0\,.
\end{eqnarray}
with
\begin{eqnarray}
b_1(r)=-\frac{\left(c_1+c_3-1\right) \left[r B'(r)-4 B(r)\right]}{2 r B(r)^3}\label{b1}\\
b_2(r)=\frac{c_1+c_3-1}{2 B(r)^2}\label{b2}\\
b_3(r)=\frac{1}{8 r^2 A(r)^3 B(r)^3}\left\{A(r) B'(r) \left[\left(c_1+c_3\right) A(r)^4 f(r)^2\right.\right.\nonumber\\\left.\left.-2 \left(c_1+c_3-2\right) A(r)^2
   f(r)+c_1+c_3\right]
   -2 B(r)\right.\nn\\\left.\times \left[\left(c_1+2 c_3-c_4\right) \left(A(r)^4 f(r)^2-1\right)\nonumber
   A'(r)+A(r)^3 f'(r)\right.\right.\nonumber\\\left.\left.\times \left(\left(c_1+2 c_3-c_4\right) A(r)^2 f(r)-3 c_1-2 c_3-c_4+4\right)\right]\right\}\label{b3}\\
b_4(r)=-\frac{1}{8 r^2 A(r)^2
   B(r)^2}\left[\left(c_1+c_3\right) A(r)^4 f(r)^2 \nonumber\right.\\\left.-2 \left(c_1+c_3-2\right) A(r)^2 f(r)+c_1+c_3\right]\,.\label{b4}
\end{eqnarray}

We will now show that $\Lambda(r,\theta)$ depends on $r$ only, and therefore equations~\eref{eqCombo} and~\eref{eq1bis} become a system
of ordinary differential equations. To show this, let us first note that solving equation~\eref{eqCombo} for $\Lambda(r,\theta)$, 
we obtain that the general structure of the solution is
\begin{equation}\label{Lambda_structure}
\Lambda(r,\theta)= \lambda_0(\theta)+\lambda_1(\theta)\Lambda_1(r)+\Lambda_2(r)\,,
\end{equation}
where $\Lambda_1(r)$  satisfies
\begin{equation}\label{lambda_1_eq}
d_3(r) \Lambda_1'(r)+d_4(r) \Lambda_1''(r)=0\,,
\end{equation}
$\Lambda_2(r)$ satisfies
\begin{equation}
d_1(r) \psi'(r)+d_2(r) \psi''(r)+
d_3(r) \Lambda_2'(r)+d_4(r) \Lambda_2''(r)=0\,,
\end{equation}
and $\lambda_0(\theta)$ and $\lambda_1(\theta)$ are generic functions (i.e. integration constants).
Because of asymptotic flatness (which prescribes that $\Lambda(r,\theta)$ asymptote to 0 
in a coordinate system that is asymptotically non-rotating), we need to set $\lambda_0(\theta)=0$.
Also, using the asymptotic spherically symmetric static solution \eref{asyF} -- \eref{asyA}
we have $d_3(r)/d_4(r) \sim D r^2$ at large radii,
with
\begin{equation}
D=\frac{64 \left(c_1-c_3-1\right) \left(c_1+c_4\right) F_1}{\left(c_1^2-c_3^2+2 c_4\right) \left(8
   A_2-3 F_1^2\right)^2}\,,
\end{equation}
so equation~\eref{lambda_1_eq} yields $\Lambda_1(r)\sim \exp(-D r^3/3)/(-D r^2)$
at large radii (once we impose that $\Lambda_1$ goes to zero at large radii because of asymptotic flatness).

Replacing then equation~\eref{Lambda_structure} into equation~\eref{eq1bis}, one obtains
\begin{equation}\label{eqCartoon}
\lambda_1''(\theta)+3 \cot \theta\lambda_1'(\theta)+k_1(r) \lambda_1(\theta) + k_2(r) =0
\end{equation}
where
\begin{eqnarray}
k_1=-\frac{2 \left\{r^4 \left[{b_3}(r) {\Lambda_1}'(r)+{b_4}(r) {\Lambda_1}''(r)\right]+{\Lambda_1}(r)\right\}}{{\Lambda_1}(r)}\\
 k_2=-\frac{2}{{\Lambda_1}(r)} \times\left\{r^4 \left[{b_1}(r) {\psi}'(r)+{b_2}(r) {\psi}''(r)\right.\right.\nonumber\\\left.\left.+{b_3}(r)
 {\Lambda_2}'(r)+{b_4}(r) {\Lambda_2}''(r)\right]+{\Lambda_2}(r)\right\}.
\end{eqnarray}
At this point, let us note that if one has functions $H(r)$, $G(\theta)$, $L(\theta)$ and $N(r)$ satifying
$H(r)+G(\theta)+N(r) L(\theta)=0$, by differentiating this equation with respect to $r$ and $\theta$, 
one immediately obtains that either $L=$ const or $N=$ const.
Applying this to equation~\eref{eqCartoon}, we find that either  $\lambda_1(\theta)=$ const, or $k_1(r)=$ const.
This latter case, however, is not possible because by replacing $\Lambda_1(r)\sim \exp(-D r^3/3)/(-D r^2)$ in the above expression for $k_1$,
and using the fact that $b_3(r) ={\cal O} (1/r^4)$ and $b_4(r) ={\cal O}(1/r^2)$ (obtained replacing equations~\eref{asyF}--\eref{asyA} in equations~\eref{b3} and~\eref{b4}),
we get $k_1\sim D^2 r^6$ at large radii. Therefore, we can conclude that the only possibility is $\lambda_1(\theta)=$ const, and from
equation~\eref{Lambda_structure} (recalling that $\lambda_0(\theta)=0$) we can conclude that $\Lambda(r,\theta)\equiv \lambda(r)$ is function of radius only,
and so is the ``frame dragging'' $\Omega(r,\theta)=2[\psi(r)+f(r) \lambda(r)/r^2]$.

 The equations that we need to solve are, therefore
 \begin{equation}\label{eeq1}
 d_1(r) \psi'(r)+d_2(r) \psi''(r)+d_3(r) \lambda'(r)+d_4(r) \lambda''(r)=0
 \end{equation}
  and
 \begin{equation}\label{eeq2}
 b_1(r) \psi'(r)+b_2(r) \psi''(r)+b_3(r) \lambda'(r)+b_4(r) \lambda''(r)+\frac{\lambda(r)}{r^4}=0\,.
 \end{equation}
While this system of equations may be integrated numerically starting from the metric horizon of the background solution
and imposing asymptotic flatness at infinity using a shooting method (c.f. the analysis in Ref.~\cite{Barausse:2011pu}), a thorough study of these solutions, their geometry 
and physical observables in the viable region of the theory's parameters space is beyond the scope of this paper, and will be presented 
elsewhere~\cite{ian_in_prep}. Here, we limit ourselves to using these equations to show that slowly rotating \ae-theory black holes 
are necessarily different than Ho\v rava gravity ones, because they present no global preferred-time slicing, and therefore no
universal horizon in the sense defined previously.

To show this, let us first note that hypersurface-orthogonality of the aether implies vanishing vorticity vector $\omega^\mu =\epsilon^{\mu\nu\alpha\beta} u_\nu \partial_\alpha u_\beta$.
Using this and the requirement that the aether be stationary and axisymmetric (hence $\partial_t u_\mu=\partial_\varphi u_\mu=0$), one obtains that it must be $u_\varphi =\ell u_t$, where
$\ell$ is a constant (of order ${\cal O}(\epsilon)$, due to the slow-rotation assumption). As noted in Ref.~\cite{letter}, a coordinate change $t'=t-\ell \varphi$
can then be shown to transform $u_\varphi$ to zero, without affecting the metric ansatz~\eref{metric}. Therefore, if one imposes hypersurface-orthogonality for the aether, it
is not restrictive to set $u_\varphi=\lambda=0$ in equations~\eref{eeq1}--\eref{eeq2}. One can then combine these equations into
a first-order equation for $\psi(r)$, namely
 \begin{eqnarray}\nn \psi'(r)
 \frac{({c_1}+{c_3}-1) ({c_1}+{c_4})}{8 r^4 A(r)^2 B(r)^4} \times\\
 \left[\left(A(r)^2 f(r)-1\right)
    A'(r)+A(r)^3 f'(r)\right]=0\,.
 \end{eqnarray}
Because the combination $\left(A(r)^2 f(r)-1\right) A'(r)+A(r)^3 f'(r)$
can be checked to be non-zero using the background solution of Ref.~\cite{Barausse:2011pu}\footnote{Alternatively
one can use the asymptotic solution \eref{asyF}--\eref{asyA} to check that this combination does not vanish identically, or
note that this combination is proportional, on the metric horizon, to the quantity $S$ defined 
in equation (21) of Ref.~\cite{letter}, which was shown to be non-zero there.},
 it follows that $\psi=\Omega=\Omega_0$ constant. However, as noted in Ref.~\cite{letter}, 
this is simply the background static spherically symmetric solution under a 
coordinate change $\varphi'=\varphi+\Omega_0 t/2$, so no slowly rotating black-hole solutions exist in \ae-theory
if the aether is required to be hypersurface-orthogonal. Therefore, no universal horizons seem to exist in \ae-theory's
slowly rotating black holes, at least in the sense defined in section~\ref{Uhor}, because no
global preferred-time foliation exists in the first place.

\subsection{Ho\v rava gravity}

As discussed in section~\ref{framework}, the Einstein equations for Ho\v rava gravity coincide with the Einstein equations
for \ae-theory, provided that those are derived by varying $g^{\mu\nu}$ and keeping $u_\mu$ fixed
and that
the aether is hypersurface-orthogonal~\cite{arxiv_note} (c.f. equations \eref{EinsteinAE} and \eref{hl1}). 
As we also stressed, the Bianchi identity implies~\cite{Jacobson:2010mx} that the  Ho\v rava gravity Einstein equations \eref{hl1} also include
equation \eref{hleq} (which follows from a variation of the action with respect to $T$), so one does not need to impose equation \eref{hleq}
explicitly when solving equations \eref{hl1}.

The only two independent fields equations at order ${\cal O}(\epsilon)$ are then given by
the $r\varphi$ and $t\varphi$ components of the Einstein equations \eref{hl1}. The former is given by equation~\eref{eq2}, and proceeding as
in \ae-theory, it implies $\omega(r,\theta)=\psi(r)$. The latter is then given by equation~\eref{eq1}, which because of
the hypersurface orthogonality condition (which as discussed above implies $\Lambda(r,\theta)=0$) becomes
\begin{equation}
\frac{{c}_{1}+c_3-1}{2 B(r)^2} \left\{\psi''(r)+\psi'(r) \left[\frac{4}{r}-\frac{B'(r)}{B(r)}\right]\right\}=0
\end{equation}
(where we have used the fact, proved in section~\ref{spin_ae}, that $\zeta_3$ and $\zeta_4$
are zero once the spherically symmetric static solution is used to evaluate them).
The solution to this equation then reads
\begin{equation}\label{rot_soln}
\Omega(r,\theta) = 2 \psi(r) = - 12 J\int_{r_{\rm H}}^r \frac{B(\rho)}{\rho^4} d\rho +\Omega_0
\end{equation}
where $r_{\rm H}$ is the location of the metric horizon, while $J$ and $\Omega_0$ are integration constants. In particular, because asymptotic flatness
imposes $B\sim 1$
far from the black hole, 
from a comparison with the Kerr solution of General Relativity, it is clear that $J$ plays the role of the black-hole spin. 
Also, it should be noted that $\Omega_0$ can be eliminated  with a coordinate transformation $\varphi'=\varphi-\Omega_0 t/2$.

Clearly, because the aether vector field does not present any terms at order ${\cal O}(\epsilon)$, the preferred-time foliation is also
unaffected at that order, and the slowly rotating solutions have the same universal horizon as the static spherically symmetric black 
holes reviewed in section~\ref{sphBH}.

\section{Conclusions}
\label{cons}

We have reviewed spherically symmmetric static black-hole solutions~\cite{Barausse:2011pu} in
\ae-theory and in the infrared limit of Ho\v rava gravity.
In order to focus on the black holes that are expected to exist in the real Universe (e.g. ones forming
from gravitational collapse), we impose that the geometry of our solutions be regular everywhere apart from the central singularity~\cite{collapse_ae}.
Because the matter fields (which couple minimally to the metric) and the propagating
modes of the gravity sector have different propagation speeds in these theories,
these black-hole solutions present multiple horizons, i.e. a ``metric'' horizon
for the matter fields, as well as spin-0, spin-1 and spin-2 horizons for the various
modes that reside in the gravity sector.
We have compared the geometry of these black holes (outside the metric horizon) to that of the Schwarzschild solution
of General Relativity, and we have discussed the prospects of constraining the differences with future electromagnetic and
gravitational-wave probes.

Lorentz-violating gravity theories are generically expected to have propagating modes that can travel arbitrarily fast in the UV limit.
This would make the very concept of event horizon (and therefore of a black hole) meaningless, because
modes with sufficiently high energies would be allowed to escape the multiple horizons mentioned above and probe the central singularity.
However, the examples we studied here demonstrate that, when there is a preferred foliation, a new kind of horizon can exist, and it acts as a causal boundary for all modes, even instantaneous ones. The nonlinear stability of universal horizon certainly requires further investigation \cite{Blas:2011ni}.

We have also investigated whether such universal horizons exist for slowly rotating black holes, and we have shown that they do indeed exist
for Ho\v rava gravity, while they do not seem to be present in \ae-theory~\cite{letter,arxiv_note}. The reason for that is that slowly rotating black holes in \ae-theory do not posses a preferred foliation. 

\section*{Acknowledgements}
 We are indebted to Ted Jacobson  
for countless stimulating conversations and insightful suggestions. Section \ref{sphBH} was largely based on results obtain in Ref.~\cite{Barausse:2011pu}, 
of which he is a coauthor.
We also thank Ian Vega for
checking the algebra of the equations of section \ref{slowly_rotating} in the context of a
related project.
E.B. acknowledges support from the European Union's Seventh Framework Programme (FP7/PEOPLE-2011-CIG)
through the Marie Curie Career Integration Grant GALFORMBHS PCIG11-GA-2012-321608. T.P.S. acknowledges financial support 
from the European Research Council under the European Union's Seventh Framework Programme (FP7/2007-2013) / ERC Grant Agreement n.~306425 ``Challenging General Relativity'' and from 
the Marie Curie Career Integration Grant LIMITSOFGR-2011-TPS Grant Agreement n.~303537.


\section*{References}
\begin{thebibliography}{100}



\bibitem{Penrose:1964wq}
  Penrose R 1965
  {\it Phys. Rev. Lett.}  {\bf 14} 57

\bibitem{Hawking:1969sw}
  Hawking S W and Penrose R 1970
  {\it Proc.\ Roy.\ Soc.\ Lond. A} {\bf 314} 529


\bibitem{ruffini} Rhoades C E and Ruffini R 1974 {\it Phys. Rev. Lett.} 32 324

\bibitem{kalogera} Kalogera V and Baym G 1996 {\it Astrophys.\ J.\ Lett.\ } 470 L61

\bibitem{darkcluster} Maoz E 1998 {\it Astrophys.\ J.\ Lett.\ } 494, L181

\bibitem{schoedel} Schoedel R {\it et al} 2002 {\it  Nature} 419 694

\bibitem{narayan_horizon} Broderick A E and Narayan R 2006
{Astrophys.\ J.\ Lett.\ } {\bf 638} L21;
Broderick A E, Loeb A and Narayan R 2009 {\it Astrophys.\ J.\ } {\bf
701} 1357


\bibitem{typeIbursts1}  
  Narayan R, Garcia M R and McClintock J E 1997 {\it  Astrophys.\ J.\ Lett. \ }  {\bf 478} L79

\bibitem{typeIbursts2}  
  Narayan R and Heyl J S 2002 {\it Astrophys.\ J.\ Lett. \ }  {\bf 574} L139

\bibitem{typeIbursts3}     
  Remillard R A, Lin D, Cooper R L and Narayan R 2006 {\it  Astrophys.\ J.\ }  {\bf 646} 407


\bibitem{kormendy} Magorrian J, Tremaine S, Richstone D {\it et al} \ 1998 {\it Astron. J.} 115 2285; Kormendy J and Richstone D \ 1995
  {\it Ann.\ Rev.\ Astron.\ Astrophys. } {\bf 33} 581


\bibitem{reverberation_mapping}  Peterson B M {\it et al.} 2004 {\it  Astrophys.\ J.\ } {\bf 613} 682;   Gebhardt K {\it et al.} 2000
  {\it Astrophys.\ J.\ Lett.\ } {\bf 543} L5



\bibitem{Kostelecky:2008ts} 
Kostelecky V A and Russell N 2011
{\it  Rev.\ Mod.\ Phys.\  }{\bf 83} 11 


\bibitem{Liberati:2013xla} 
Liberati S 2013
 {\it  Class.\ Quant.\ Grav.\ }{\bf 30} 133001

\bibitem{Jacobson:2000xp} 
  Jacobson T and Mattingly D 2001
  {\it Phys.\ Rev.\ D} {\bf 64} 024028
  
\bibitem{arXiv:0901.3775} 
  Ho\v rava P 2009 
  {\it Phys.\ Rev.\ D\ } {\bf 79} 084008
  
\bibitem{Jacobson:2010mx} 
  Jacobson T 2010
  {\it Phys.\ Rev.\ D}  {\bf 81} 101502
  [Erratum-ibid.\ 2010 {\bf 82} 129901]
  

 \bibitem{Barausse:2011pu} 
   Barausse E, Jacobson T and Sotiriou T P 2011
   {\it  Phys.\ Rev.\ D } {\bf 83} 124043 


 \bibitem{letter} Barausse E and Sotiriou T P 2012,
  {\it  Phys.\ Rev.\ Lett.\ } {\bf 109} 181101 
  [Erratum-ibid.\ 2013 {\bf 110} 039902(E)].

   \bibitem{arxiv_note}   Barausse E and Sotiriou E 2013
 {\it  Phys.\ Rev.\ D } {\bf 87} 087504

   
\bibitem{Blas:2009qj}
  Blas D, Pujolas O and Sibiryakov S 2010,
 {\it Phys.\ Rev.\ Lett.\ } {\bf 104} 181302
 
\bibitem{Blas:2009yd} 
  Blas D, Pujolas O and Sibiryakov S 2009,
 {\it  JHEP} {\bf 0910} 029
 

\bibitem{Sotiriou:2009bx}
  Sotiriou T P, Visser M and Weinfurtner S 2009
  {\it Phys.\ Rev.\ Lett.\ } {\bf 102} 251601;
  2009 {\it JHEP} {\bf 0910} 033
  [arXiv:0905.2798];
  
\bibitem{Weinfurtner:2010hz} 
  Weinfurtner S, Sotiriou T P and Visser M 2010
  {\it J.\ Phys.\ Conf.\ Ser.\ } {\bf 222} 012054 
  
\bibitem{Horava:2010zj} 
  Ho\v rava P and Melby-Thompson C M 2010
  {\it Phys.\ Rev.\ D} {\bf 82} 064027 
  
\bibitem{Vernieri:2011aa} 
  Vernieri D and Sotiriou T P 2012
  {\it Phys.\ Rev.\ D} {\bf 85}, 064003 
  
  
\bibitem{Vernieri:2012ms} 
  Vernieri D and Sotiriou T P 2012
  arXiv:1212.4402 [hep-th].
  
\bibitem{Sotiriou:2010wn} 
  Sotiriou T P 2011
  {\it J.\ Phys.\ Conf.\ Ser.\ } {\bf 283} 012034 
  
\bibitem{Eling:2006ec}
  Eling C and Jacobson J 2006
  {\it Class.\ Quant.\ Grav.\ } {\bf 23} 5643
  
\bibitem{Burgess:2002tb} 
  Burgess C P, Cline J M, Filotas E, Matias J and Moore G D 2002
  {\it JHEP} {\bf 0203} 043
  
\bibitem{Iengo:2009ix} 
  Iengo R, Russo J G and Serone M 2009
  {\it JHEP} {\bf 0911} 020
  
\bibitem{Pospelov:2010mp} 
  Pospelov M and Shang Y 2012
  {\it Phys.\ Rev.\ D} {\bf 85} 105001
  
\bibitem{Blas:2010hb} 
  Blas D, Pujolas O and Sibiryakov S 2011
  {\it JHEP} {\bf 1104} 018
  
  
\bibitem{Liberati:2012jf} 
  Liberati S, Maccione L and Sotiriou T P 2012
  {\it Phys.\ Rev.\ Lett.\ } {\bf 109} 151602
  
\bibitem{Papazoglou:2009fj} 
  Papazoglou A and Sotiriou T P 2010
  {\it Phys.\ Lett.\ B } {\bf 685} 197
  
\bibitem{Kimpton:2010xi} 
  Kimpton I and Padilla A 2010
  {\it JHEP} {\bf 1007} 014
  
\bibitem{Blas:2009ck} 
  Blas D, Pujolas O and Sibiryakov S 2010
  {\it Phys.\ Lett.\ B} {\bf 688} 350 

\bibitem{Will:2005va}
  Will C M 2006
  {\it Living Rev.\ Rel.\ }  {\bf 9} 3 
    

\bibitem{Eling:2003rd}
  Eling C and Jacobson T 2004
 {\it Phys.\ Rev.\ D} {\bf 69} 064005
  
\bibitem{Elliott:2005va}
  Elliott J W, Moore G D and Stoica H
  {\it JHEP} {\bf 0508} 066.
  
\bibitem{Foster:2007gr}
  Foster B Z 2007
  {\it Phys.\ Rev.\  D} {\bf 76} 084033
  
\bibitem{Foster:2006az}
  Foster B Z 2006
  {\it Phys.\ Rev.\  D} {\bf 73} 104012
  [Erratum-ibid.\  2007 {\bf 75} 129904]
  
\bibitem{Blas:2011zd} 
  Blas D and Sanctuary H 2011
  {\it Phys.\ Rev.\ D } {\bf 84} 064004 
  
\bibitem{sensitivities_AE} Yagi K, Blas D, Yunes N and Barausse E 2013 arXiv:1307.6219
\bibitem{Zuntz:2008zz} 
  Zuntz J A, Ferreira P G and Zlosnik T G 2008,
  {\it Phys.\ Rev.\ Lett.\ } {\bf 101} 261102
  
  
\bibitem{Audren:2013dwa} 
  Audren B, Blas D, Lesgourgues J and Sibiryakov S 2013
  arXiv:1305.0009 [astro-ph.CO].


\bibitem{Blas:2011ni} 
  Blas D and Sibiryakov S 2011
  {\it Phys.\ Rev.\ D } {\bf 84} 124043

\bibitem{Gn_vs_Gbare} Carroll S M and Lim E A {\it Phys. Rev. D } 2004 {\bf 70} 123525
\bibitem{diffeo_ted}   Jacobson T 2011 {\it Class.\ Quant.\ Grav.\ } {\bf 28} 245011

\bibitem{singularBHs} Tamaki T and Miyamoto U 2008 {\it  Phys.\ Rev.\  D} 77 024026 
\bibitem{collapse_ae} Garfinkle D, Eling C and Jacobson T 2007 {\it  Phys.\ Rev.\  D} {\bf 76} 024003
\bibitem{NS_ae} Eling C and Jacobson T 2006  {\it Class.\ Quant.\ Grav.\ } {\bf 23} 5625   [Erratum-ibid.\ 2010  {\bf 27} 049801]

\bibitem{thindisks}   Shakura N I and Sunyaev R A 1973 {\it Astron.\ Astrophys.\ } {\bf 24} 337; Novikov I D and Thorne K S 1973 in Dewitt C and Dewitt B S ed., 
{\it Black Holes (Les Astres Occlus) Astrophysics of black holes},  pp 343–450

\bibitem{continuum} McClintock J, Narayan R, Davis S, {\it et al.\ } 2011 {\it  Class.\ Quant.\ Grav.\ } {\bf 28} 114009
\bibitem{iron} Miller J M, Reynolds C S, Fabian A C, Miniutti G, and Gallo L C 2009 {\it Astrophys.\ J.} {\bf 697} 900 

\bibitem{psaltis_iron} Johannsen T and Psaltis D 2012 {\it Astrophys.\ J.\ } in press, arXiv:1202.6069 
\bibitem{BB10} Bambi C and Barausse E 2011 {\it  Astrophys.\ J.\ }  {\bf 731} 121
\bibitem{BB11} Bambi C and Barausse E 2011 {\it Phys.\ Rev.\ D } {\bf 84} 084034
\bibitem{shadows1} Johannsen T and Psaltis D 2010 {\it  Astrophys.\ J.\ } {\bf 718} 446
\bibitem{shadows2} Bambi C and Freese K {\it  Phys.\ Rev.\ D } {\bf 79} 043002
\bibitem{elisa}  Seoane P A{\it et al.}  [the eLISA Collaboration] 2013 arXiv:1305.5720 
  
 \bibitem{sergey} The concept of a universal horizon in this context was first introduced by Sergey Sibiryakov in the 2010 Peyresq meeting.
  
\bibitem{Dubovsky:2006vk} 
  Dubovsky S L and Sibiryakov S 2006
  {\it Phys.\ Lett.\ B } {\bf 638} 509 
  
\bibitem{Eling:2007qd} 
  Eling C, Foster B Z, Jacobson T and Wall A C 2007
  {\it Phys.\ Rev.\ D } {\bf 75} 101502
  
\bibitem{Berglund:2012bu} 
  Berglund P, Bhattacharyya J and Mattingly D 2012
  {\it Phys.\ Rev.\ D } {\bf 85} 124019
  
\bibitem{Berglund:2012fk} 
  Berglund P, Bhattacharyya J and Mattingly D 2013
  {\it Phys.\ Rev.\ Lett.\ } {\bf 110} 071301
  
\bibitem{Cropp:2013zxi} 
  Cropp B, Liberati S and Visser M 2013
  arXiv:1302.2383 [gr-qc].




 \bibitem{natal} Liu L, McClintock J, Narayan R, Davis S and Orosz J 2008
  {\it Astrophys.\ J.\ Lett. } {\bf 679} L37; 
Shafee R, McClintock J, Narayan R, {\it et al. } \ 2006  {\it Astrophys.\ J.\ Lett.\ } {\bf 636} L113;
McClintock J, Shafee R, Narayan R, Remillard R A, Davis S and Li L 2006 {\it Astrophys.\ J.\ }  {\bf 652} 518.

\bibitem{iron_smbh} Brenneman L W and Reynolds  C S 2009 {\it Astrophys.\ J.} {\bf 702} 1367 

\bibitem{soltan_argument}   Li Y R, Wang J M and Ho L C 2012, {\it  Astrophys.\ J. }  {\bf 749} 187; Shankar F, Weinberg D H and Miralda-Escud{\'e} J 2013 {\it Mon.\ Not.\ Roy.\ Astron.\ Soc.\ } {\bf 428} 421

\bibitem{spin_evolution} Berti E and Volonteri M 2008  {\it Astrophys.\ J.\ }
{\bf 684} 822;
Fanidakis N \textit{et al.} 2011 {\it Mon.\ Not.\ Roy.\ Astron.\ Soc.\ } {\bf
410} 53;  Barausse E 2012 {\it Mon.\ Not.\ Roy.\ Astron.\ Soc.\ } {\bf
423} 2533


\bibitem{BZeffect} Blandford R D and Znajek R L 1977  {\it Mon.\ Not.\ Roy.\ Astron.\ Soc.\ } {\bf 179} 433 

\bibitem{hartle_thorne} Hartle J B 1967
  {\it Astrophys.\ J.\ } {\bf 150} 1005;  Hartle J B and Thorne K S 1968
  {\it  Astrophys.\ J.\ } {\bf 153} 807 
%
\bibitem{ian_in_prep} Vega I, Barausse E and Sotiriou T P 2013 {\it in preparation}

\end{thebibliography}
\end{document}